\documentclass[journal,twocolumn]{IEEEtran}
\usepackage{graphicx}
\usepackage{url}
\usepackage{booktabs}
\usepackage{enumerate}
\usepackage{hyperref}
\usepackage{lipsum}
\usepackage{acronym}
\usepackage{subcaption}
\usepackage{float}
\usepackage{cite}
\usepackage[export]{adjustbox}
\usepackage{color,soul} 

\begin{document}

\title{Unsupervised Machine Learning for Networking: \\Techniques, Applications and Research Challenges}
	\author{Muhammad Usama$^{1}$, Junaid Qadir$^{1}$, Aunn Raza$^{2}$, Hunain Arif$^{2}$, Kok-Lim Alvin Yau$^{3}$,\\ Yehia Elkhatib$^{4}$, Amir Hussain$^{5}$, Ala Al-Fuqaha$^{6}$\\	
\normalsize $^{1}$Information Technology University (ITU)-Punjab, Lahore, Pakistan\\
\normalsize $^{2}$School of EE and CS, National University of Sciences and Technology (NUST), Pakistan\\
\normalsize $^{2}$Sunway University, Malaysia\\	
\normalsize $^{4}$MetaLab, School of Computing and Communications, Lancaster University, UK\\
\normalsize $^{5}$University of Stirling, United Kingdom\\	
\normalsize $^{6}$Western Michigan University, United States of America\\	
		}
	
	\date{}
	\maketitle

\begin{abstract}
While machine learning and artificial intelligence have long been applied in networking research, the bulk of such works has focused on supervised learning. Recently there has been a rising trend of employing unsupervised machine learning using unstructured raw network data to improve network performance and provide services such as traffic engineering, anomaly detection, Internet traffic classification, and quality of service optimization. The interest in applying unsupervised learning techniques in networking emerges from their great success in other fields such as computer vision, natural language processing, speech recognition, and optimal control (e.g., for developing autonomous self-driving cars). Unsupervised learning is interesting since it can unconstrain us from the need of labeled data and manual handcrafted feature engineering thereby facilitating flexible, general, and automated methods of machine learning. The focus of this survey paper is to provide an overview of the applications of unsupervised learning in the domain of networking. We provide a comprehensive survey highlighting the recent advancements in unsupervised learning techniques and describe their applications for various learning tasks in the context of networking. We also provide a discussion on future directions and open research issues, while also identifying potential pitfalls. While a few survey papers focusing on the applications of machine learning in networking have previously been published, a survey of similar scope and breadth is missing in literature. Through this paper, we advance the state of knowledge by carefully synthesizing the insights from these survey papers while also providing contemporary coverage of recent advances.

\end{abstract}
		
\section{Introduction}

Networks---such as the Internet and mobile telecom networks---serve the function of the central hub of modern human societies, which the various threads of modern life weave around. With networks becoming increasingly dynamic, heterogeneous, and complex, the management of such networks has become less amenable to manual administration, and can benefit from leveraging support from methods for optimization and automated decision-making from the fields of artificial intelligence (AI) and machine learning (ML). Such AI and ML techniques have already transformed multiple fields---e.g., computer vision, natural language processing (NLP), speech recognition, and optimal control (e.g., for developing autonomous self-driving vehicles)---with the success of these techniques mainly attributed to \textit{firstly}, significant advances in unsupervised ML techniques such as deep learning, \textit{secondly}, the ready availability of large amounts of unstructured raw data amenable to processing by unsupervised learning algorithms, and \textit{finally}, advances in computing technologies through advances such as cloud computing, graphics processing unit (GPU) technology and other hardware enhancements. It is anticipated that AI and ML will also make a similar impact on the networking ecosystem and will help realize a future vision of \textit{cognitive networks} \cite{thomas2007cognitive} \cite{latif2017artificial}, in which networks will self-organize and will autonomously implement intelligent network-wide behavior to solve problems such as routing, scheduling, resource allocation, and anomaly detection. 

\begin{table*}[!ht]
\centering
\scriptsize
\begin{tabular}{@{}|p{2.2cm}|p{2cm}|p{.7cm}|p{1cm}|p{3.5cm}|p{1.5cm}|p{1cm}|p{1cm}|p{1.5cm}|@{}}
\hline
Survey paper & Published In & Year& \# References & Areas Focused &Unsupervised ML &Deep Learning &Pitfalls&Future Challenges\\ \hline 

Patcha et al.\cite{patcha2007overview} & Elsevier\newline Computer Networks &2007&100&ML for Network Intrusion Detection&$\approx$&$\times$&$\times$& $\surd$ \\ \hline

Nguyen et al.\cite{nguyen2008survey} & IEEE COMST& 2008&68& ML for Internet Traffic Classification &$\approx$&$\times$&$\times$&$\times$ \\ \hline 

Bkassiny et al.\cite{bkassiny2013survey} & IEEE COMST &2013&177&ML for Cognitive Radios&$\approx$&$\times$&$\times$&$\times$ \\ \hline

Alsheikh et al. \cite{alsheikh2014machine} & IEEE COMST&2015&152&ML for WSNs &$\approx$&$\times$&$\times$&$\surd$ \\ \hline

Buczak et al.\cite{Buczak2016} & IEEE COMST &2016&113&ML for Cyber Security Intrusion Detection&$\approx$&$\times$&$\times$&$\surd$\\ \hline 

Klaine et al. \cite{klaine2017survey} & IEEE COMST&2017&269&ML in SONs &$\approx$&$\times$&$\times$&$\surd$ \\ \hline

Meshram et al. \cite{meshram2017anomaly} & Springer\newline Book Chapter &2017&16&ML for Anomaly Detection in Industrial Networks& $\approx$&$\times$&$\times$&$\surd$\\ \hline 

Fadlullah et al.\cite{fadlullah2017state} & IEEE COMST &2017&260&ML for Network Traffic Control&$\approx$&$\surd$&$\times$&$\surd$ \\ \hline

Hodo et al. \cite{hodo2017shallow} & ArXiv &2017&154&ML Network Intrusion Detection& $\approx$&$\surd$&$\times$&$\times$\\ \hline 

This Paper& - & 2017 & 323 & Unsupervised ML in Networking &$\surd$&$\surd$&$\surd$&$\surd$ \\ \hline


\end{tabular}
\caption{Comparison of our paper with existing survey and review papers. (Legend: \protect$\surd$ means covered; \protect$\times$ means not covered; \protect$\approx$ means partially covered.)}
\label{surveyTable}
\end{table*}

    \begin{table}[!ht]
		\caption{List of common acronyms used \label{tab:acronymsused}}
		\centering
		\scriptsize
		\begin{tabular}{cl}
			\toprule
    ADS & Anomaly Detection System\\
    A-NIDS & Anomaly \& Network Intrusion Detection System\\
    AI & Artificial Intelligence\\
    ANN & Artificial Neural Network\\
    ART & Adaptive Resonance Theory\\
    BSS & Blind Signal Separation\\
    BIRCH & Balanced Iterative Reducing and Clustering Using Hierarchies\\
    CDBN & Convolutional Deep Belief Network\\
    CNN & Convolutional Neural Network\\
    CRN & Cognitive Radio Network\\
    DBN & Deep Belief Network\\
    DDoS & Distributed Denial of Service\\
    DNN & Deep Neural Network\\
    DNS & Domain Name Service\\
    DPI & Deep Packet Inspection\\
    EM & Expectation-Maximization\\
    GTM & Generative Topographic Model\\
    GPU & Graphics Processing Unit\\
    GMM & Gaussian Mixture Model \\
    HMM & Hidden Markov Model\\
    ICA & Independent Component Analysis\\
    IDS & Intrusion Detection System\\
    IoT & Internet of Things\\
    LSTM & Long Short-Term Memory\\
    LLE & Locally Linear Embedding\\
    LRD & Low Range Dependencies\\
    MARL & Multi-Agent Reinforcement Learning\\
    ML & Machine Learning\\
    MLP & Multi-Layer Perceptron\\
    MRL & Model-based Reinforcement Learning\\
    MDS & Multi-Dimensional Scaling\\
    MCA & Minor Component Analysis\\
    NMF & Non-Negative Matrix Factorization\\
    NMS & Network Management System\\
    NN & Neural Network\\
    NMDS & Nonlinear Multi-dimensional Scaling\\
    OSPF & Open Shortest Path First\\
    PU & Primary User\\
    PCA & Principal Component Analysis\\
    PGM & Probabilistic Graph Model\\
    QoE & Quality of Experience\\
    QoS & Quality of Service\\
    RBM & Restricted Boltzmann Machine\\
    RL & Reinforcement Learning\\
    RLFA & Reinforcement Learning with Function Approximation\\
    RNN & Recurrent Neural Network\\
    SDN & Software Defined Network\\
    SOM & Self-Organizing Map\\
    SON & Self-Organizing Network\\
    SVM & Support Vector Machine\\
    SON & Self Organizing Network\\
    SSAE & Shrinking Sparse Autoencoder\\
    TCP & Transmission Control Protocol\\
    t-SNE & t-Distributed Stochastic Neighbor Embedding\\
    TL & Transfer Learning\\
    VoIP & Voice over IP\\
    VoQS & Variation of Quality Signature\\
    VAE & Variational Autoencoder \\
    WSN & Wireless Sensor Network\\
    
			\bottomrule
		\end{tabular}
	\end{table}

\begin{figure*}
    	\begin{center}
    	 		\includegraphics[width=1.0\textwidth]{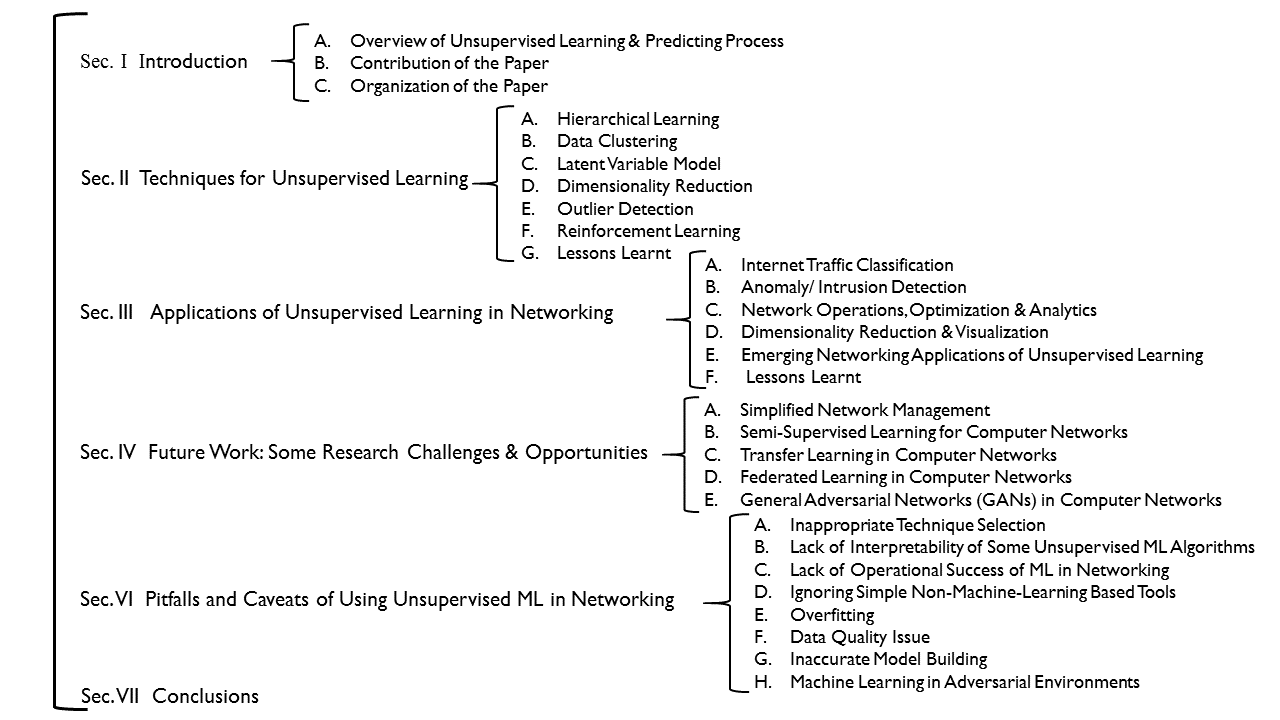}
    		 \caption{Outline of the Paper}
    		\label{fig:outline}
     	\end{center}
     \end{figure*}

     The initial attempts towards creating cognitive or intelligent networks have relied mostly on \textit{supervised ML methods}, which are efficient and powerful, but are limited in scope by their need for labeled data. With network data becoming increasingly voluminous (with a disproportionate rise in unstructured unlabeled data), there is a groundswell of interest in leveraging \textit{unsupervised ML methods} to utilize unlabeled data, in addition to labeled data where available, to optimize network performance \cite{qadir2015ieee}. The rising interest in applying unsupervised ML in networking applications also stems from the need to liberate ML applications from restrictive demands of supervised ML for labeled networking data, which is expensive to curate at scale (since labeled data may be unavailable and manual annotation prohibitively inconvenient) in addition to being suspect to being outdated quickly (due to the highly dynamic nature of computer networks) \cite{suthaharan2014big}. 

We are already witnessing the failure of human network administrators to manage and monitor all bits and pieces of network \cite{shenker2011future}, and the problem will only exacerbate with further growth in the size of networks with paradigms such as the Internet of things (IoT). An ML-based network management system (NMS) is desirable in such large networks so that faults/bottlenecks/anomalies may be predicted in advance with reasonable accuracy. In this regard, networks already have ample amount of untapped data, which can provide us with decision-making insights making networks more efficient and self-adapting. With unsupervised ML, the pipe dream is that every algorithm for adjusting network parameters (be it, TCP congestion window or rerouting network traffic in peak time) will optimize itself in a self-organizing fashion according to the environment and application, user, and network's Quality of Service (QoS)  requirements and constraints \cite{malik2015qos}. Unsupervised ML methods, in concert with existing supervised ML methods, can provide a more efficient method that lets a network manage, monitor, and optimize itself, while keeping the human administrators in the loop with the provisioning of timely actionable information. 


Unsupervised ML techniques facilitate the analysis of raw datasets, thereby helping in generating analytic insights from unlabeled data. Recent advances in hierarchical learning, clustering algorithms, factor analysis, latent models, and outlier detection, have helped significantly advance the state of the art in unsupervised ML techniques. Unsupervised ML has many applications such as feature learning, data clustering, dimensionality reduction, anomaly detection, etc. In particular, recent unsupervised ML advances---such as the development of ``deep learning'' techniques \cite{lecun2015deep}---have however significantly advanced the ML state of the art by facilitating the processing of raw data without requiring careful engineering and domain expertise for feature crafting. The versatility of deep learning and distributed ML can be seen in the diversity of their applications that range from self-driving cars to the reconstruction of brain circuits \cite{lecun2015deep}. Unsupervised learning is also often used in conjunction with supervised learning in a \textit{semi-supervised learning} setting to preprocess the data before analysis and thereby help in crafting a good feature representation and in finding patterns and structures in unlabeled data. 




The rapid advances in deep neural networks, the democratization of enormous computing capabilities through cloud computing and distributed computing, and the ability to store and process large swathes of data, have motivated a surging interest in applying unsupervised ML techniques in the networking field. The field of networking also appears to be well suited to, and amenable to applications of unsupervised ML techniques, due to the largely distributed decision-making nature of its protocols, the availability of large amounts of network data, and the urgent need for \textit{intelligent/cognitive} networking. Consider the case of routing in networks. Networks these days have evolved to be very complex, and they incorporate multiple physical paths for redundancy and utilize complex routing methodologies to direct the traffic. Our application traffic does not always take the optimal path we would expect, leading to unexpected and inefficient routing performance. To tame such complexity, unsupervised ML techniques can autonomously self-organize the network taking into account a number of factors such as real-time network congestion statistics as well as application QoS requirements \cite{qadir2016artificial}.

The purpose of this paper is to highlight the important advances in unsupervised learning, and after providing a tutorial introduction to these techniques, to review how such techniques have been, or could be, used for various tasks in modern next-generation networks comprising both computer networks as well as mobile telecom networks.

\textit{Contribution of the paper:} To the best of our knowledge, there does not exist a survey that specifically focuses on the important applications of unsupervised ML techniques in networks, even though a number of surveys exist that focus on specific ML applications pertaining to networking---for instance, surveys on using ML for cognitive radios \cite{bkassiny2013survey}, traffic identification and classification \cite{nguyen2008survey}, anomaly detection \cite{patcha2007overview} \cite{meshram2017anomaly}. Previous survey papers have either focused on specific unsupervised learning techniques (e.g., Ahad et al. \cite{ahad2016neural} provided a survey of the applications of neural networks in wireless networks) or on some specific applications of computer networking (Buczak and Guven \cite{Buczak2016} have provided a survey of the applications of ML in cyber intrusion detection). Our survey paper is timely since there is great interest in deploying automated and self-taught unsupervised learning models in the industry and academia. Due to relatively limited applications of unsupervised learning in networking---in particular, the deep learning trend has not yet impacted networking in a major way---unsupervised learning techniques hold a lot of promises for advancing the state of the art in networking in terms of adaptability, flexibility, and efficiency. The novelty of this survey is that it covers many different important applications of unsupervised ML techniques in computer networks and provides readers with a comprehensive discussion of the unsupervised ML trends, as well as the suitability of various unsupervised ML techniques. A tabulated comparison of our paper with other existing survey and review articles is presented in Table \ref{surveyTable}.

\textit{Organization of the paper:} The organization of this paper is depicted in Figure \ref{fig:outline}. Section \ref{sec:technique} provides a discussion on various unsupervised ML \textit{techniques} (namely, hierarchical learning, data clustering, latent variable models, outlier detection and reinforcement learning). Section \ref{sec:wwc} presents a survey of the \textit{applications} of unsupervised ML specifically in the domain of computer networks. Section \ref{futurework} describes future work and opportunities with respect to the use of unsupervised ML in future networking. Section \ref{pitfalls} discusses a few major pitfalls of the unsupervised ML approach and its models. Finally, Section \ref{conclusions} concludes this paper. For the reader's facilitation, Table \ref{tab:acronymsused} shows all the acronyms used in this survey for convenient referencing. 


\section{Techniques for Unsupervised Learning}
\label{sec:technique}
	
\begin{figure*}
        	\centering
        		\includegraphics[width=.68\textwidth, center]{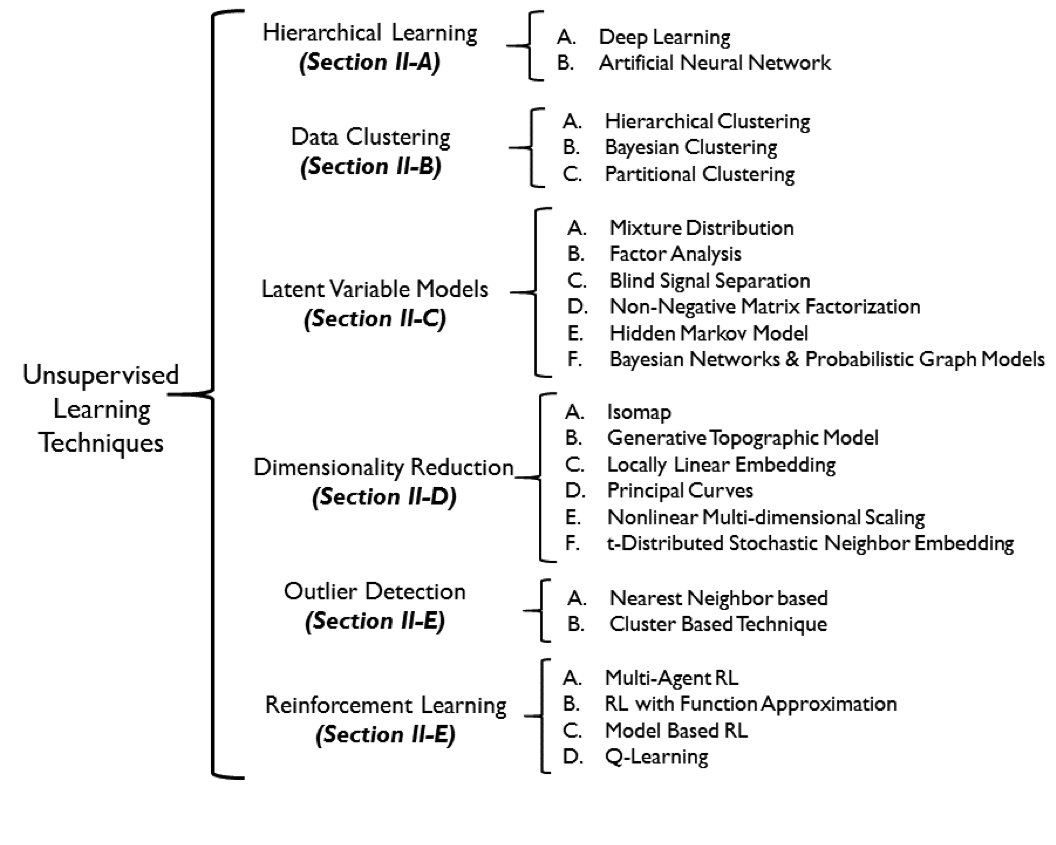}
        		\caption{Taxonomy of Unsupervised Learning Techniques}
        		\label{fig:tult}
\end{figure*}

In this section, we will introduce some widely used unsupervised learning techniques and their applications in computer networks. We have divided unsupervised learning techniques into five major categories: hierarchical learning, data clustering, latent variable models, outlier detection, and reinforcement learning. Figure \ref{fig:tult} depicts a taxonomy of unsupervised learning techniques and also notes the relevant sections in which these techniques are discussed.

\subsection{Hierarchical Learning}

Hierarchical learning is defined as learning simple and complex features from a hierarchy of multiple linear and nonlinear activations. In learning models, a feature is a measurable property of the input data. Desired features are ideally informative, discriminative, and independent. In statistics, features are also known as explanatory (or independent) variables \cite{guyon2008feature}. Feature learning (also known as data representation learning) is a set of techniques that can learn one or more features from input data \cite{coates2011analysis}. It involves the transformation of raw data into a quantifiable and comparable representation, which is specific to the property of the input but general enough for comparison to similar inputs. Conventionally, features are handcrafted specific to the application on hand. It relies on domain knowledge but even then they do not generalize well to the variation of real world data, which gives rise to automated learning of generalized features from the underlying structure of the input data. Like other learning algorithms, feature learning is also divided among domains of supervised and unsupervised learning depending on the type of available data. Almost all unsupervised learning algorithms undergo a stage of feature extraction in order to learn data representation from unlabeled data and generate a feature vector on the basis of which further tasks are performed.
	    
\begin{figure*}
        	\begin{center}
        		\includegraphics[width=.8\textwidth]{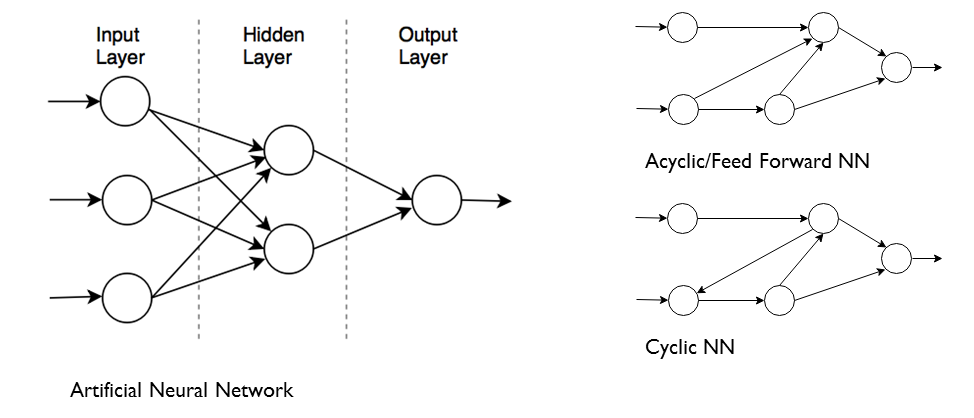}
        		\caption{Illustration of an ANN (Left); Different types of ANN topologies (Right)}  
        		\label{fig:ann}
        	\end{center}
        \end{figure*}

\begin{figure*}
        	\begin{center}
        		\includegraphics[width=.8\textwidth]{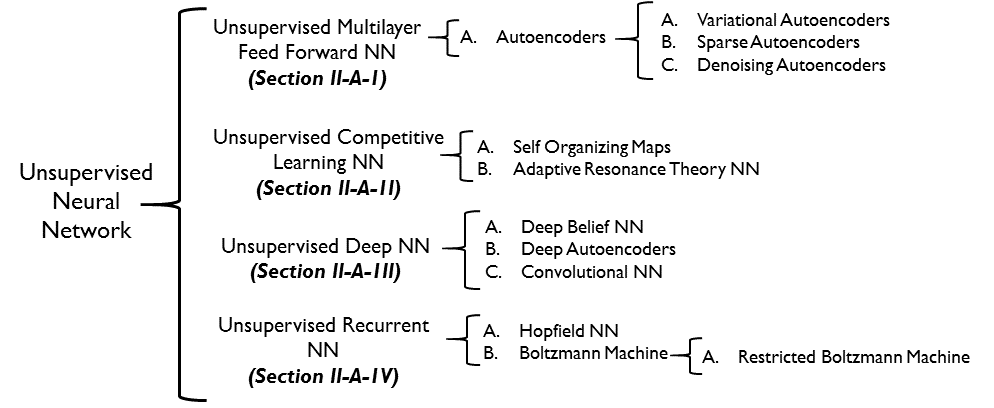}
        		\caption{Taxonomy of Unsupervised Neural Networks}
        		\label{fig:NN_Taxonomy}
        	\end{center}
        \end{figure*}

Hierarchical learning is intimately related to two strongly correlated areas: deep learning and neural networks. In particular, deep learning techniques benefits from the fundamental concept of artificial neural networks (ANNs), a deep structure consists of multiple hidden layers with multiple neurons in each layer, a nonlinear activation function, a cost function and a back-propagation algorithm. Deep learning \cite{goodfellow2016deep} is a hierarchical technique that models high level abstraction in data using many layers of linear and nonlinear transformations. With deep enough stack of these transformation layers, a machine can self-learn a very complex model or representation of data. Learning takes place in hidden layers and the optimal weights and biases of the neurons are updated in two passes, namely, feed forward and back-propagation. A typical ANN and typical cyclic and acyclic topologies of interconnection between neurons are shown in Figure \ref{fig:ann}. A brief taxonomy of Unsupervised NNs is presented in Figure \ref{fig:NN_Taxonomy}.

   \begin{table*}
	\caption{Applications of Hierarchical Learning/ Deep Learning in Networking Applications \label{HLDL}}
    {
	\centering
		\scriptsize
	\begin{tabular}{ p{5cm} p{2cm}  p{10cm}  }
		
		\toprule
		Reference &  Technique & Brief Summary\\
		\midrule
		
		\textit{\textbf{\underline{Internet Traffic Classification}}} \\
		\\
       
    Lotfollahi et al. \cite{deeppacket}&SAE \& CNN& SAE and CNN were used for feature extraction from the Internet traffic data for classification and characterizing purpose.\\
        
    Wang et al. \cite{wang2017malware}&CNN& CNN is used to extract features from the Internet traffic where traffic is considered as an image for malware detection. \\
        
    Yousefi et al.\cite{yousefi2017autoencoder}&Autoencoder&Autoencoder is used as a generative model to learn the latent feature representation of network traffic vector, for cyber attack detection and classification. \\
		
		\midrule
		\textit{\textbf{\underline{Anomaly/Intrusion Detection}}} \\
		\\
	    
    Aygun et al.\cite{aygun2017network}&Denoising Autoencoder&Stochastically Improved autoencoder and denosing autoencoder are used to learn feature for zero day anomaly detection in Internet traffic. \\
	    
	Putchala et al.\cite{putchala2017deep}&RNN& Gated recurrent unit and random forest techniques are used for feature extraction and anomaly detection in IoT data. \\
	    
	Tuor et al.\cite{tuor2017deep}&RNN& RNN and DNN are employed to extract feature from raw data which then used for threat assessment and insider threat detection in data streams. \\
		
		\midrule
	\textit{\textbf{\underline{Network Operations, Optimization and Analytics}}} \\
	\\

	Aguiar et al.\cite{aguiar2012real}&Random Neural Network& Random neural network are used for extracting the quality behavior of multimedia application for improving the QoE of multimedia applications in wireless mesh network. \\
		
	Piamrat et al.\cite{piamrat2008qoe}&Random Neural Network& Random neural network are used for learning the mapping between QoE score and technical parameters so that it can give QoE score in real-time for multimedia applications in IEEE 802.11 wireless networks. \\
		
		\midrule
	\multicolumn{3}{l}{\textit{\textbf{\underline{Emerging Networking Application of Unsupervised Learning}}}} \\
	\\
	
	Karra et al.\cite{karra2017modulation}&DNN\&CNN&Hierarchical learning is used for feature extraction from spectrogram snap shots of signal for modulation detection in communication system based on software defined radio. \\
	
	Zhang et al.\cite{zhang2017convolutional}&CNN&Convolutional filters are used for feature extraction from cognitive radio waveforms for automatic recognition. \\
	
	Moysen et al.\cite{moysen20174g}&ANN&Authors expressed ANN as a recommended system to learn the hierarchy of the output, which is later used in SON. \\
	
	Xie et al.\cite{xie2017iot}&RNN&RNN variant LSTM is used for learning memory based hierarchy of time interval based IoT sensor data, from smart cities datasets. \\
		 	
		   \hline
		\end{tabular}
	}
\end{table*}

An ANN has three types of layers (namely input, hidden and output, each having different activation parameters). \textit{Learning} is the process of assigning optimal activation parameters enabling ANN to perform input to output mapping. For a given problem, an ANN may require multiple hidden layers involving long chain of computations, i.e., its \textit{depth} \cite{schmidhuber2015deep}. Deep learning has revolutionized ML and is now increasingly being used in diverse settings---e.g., object identification in images, speech transcription into text, matching user's interests with items (such as news items, movies, products) and making recommendations, etc. But until 2006, relatively few people were interested in deep learning due to the high computational cost of deep learning procedures. It was widely believed that training deep learning architectures in an unsupervised manner was intractable, and supervised training of deep NNs (DNN) also showed poor performance with large generalization errors \cite{bengio2009learning}. However, recent advances \cite{hinton2006fast,bengio2007greedy,poultney2006efficient} have shown that deep learning can be performed efficiently by separate unsupervised pre-training of each layer with the results revolutionizing the field of ML. Starting from the input (observation) layer, which acts as an input to the subsequent layers, pre-training tends to learn data distributions while the usual supervised stage performs local search for fine-tuning. 


        
\vspace{2mm}        
\subsubsection{Unsupervised Multilayer Feed Forward NN}

Unsupervised multilayer feed forward NN, with reference to graph theory, has a directed graph topology as shown in Figure \ref{fig:ann}. It consists of no cycles, i.e., does not have feedback path in input propagation through NN. Such kind of NN is often used to approximate a nonlinear mapping between inputs and required outputs. Autoencoders are the prime examples of unsupervised multilayer feed forward NNs. 

\paragraph{Autoencoders}

An autoencoder is an unsupervised learning algorithm for ANN used to learn compressed and encoded representation of data, mostly for dimensionality reduction and for unsupervised pre-training of feed forward NNs. Autoencoders are generally designed using approximation function and trained using backpropagation and stochastic gradient decent (SGD) techniques. Autoencoders are the first of their kind to use back-propagation algorithm to train with unlabeled data. Autoencoders aim to learn compact representation of the function of input using the same number of input and output units with usually less hidden units to encode a feature vector. They learn the input data function by recreating the input at the output, which is called encoding/decoding, to learn at the time of training NN.  In short, a simple autoencoder learns low-dimensional representation of the input data by exploiting similar recurring patterns.

Autoencoders have different variants \cite{ngiam2011optimization} such as variational autoencoders, sparse autoencoders, and denoising autoencoders. \textit{Variational autoencoder} is an unsupervised learning technique used clustering, dimensionality reduction and visualization, and for learning complex distributions \cite{doersch2016tutorial}. In a\textit{ sparse autoencoder}, a sparse penalty on the latent layer is applied for extracting unique statistical feature from unlabeled data. Finally, \textit{denoising autoencoders} are used to learn the mapping of a corrupted data point to its original location in the data space in unsupervised manner for manifold learning and reconstruction distribution learning.

%

\vspace{2mm}
\subsubsection{Unsupervised Competitive Learning NN}
\label{competitiveNNs}

Unsupervised competitive learning NNs is a winner-take-all neuron scheme, where each neuron competes for the right of the response to a subset of the input data. This scheme is used to remove the redundancies from the unstructured data. Two major techniques of unsupervised competitive learning NNs are self-organizing maps and adaptive resonance theory NNs. 

\vspace{2mm}
\textit{Self-Organizing/ Kohonen Maps}: Self-Organizing Maps (SOM), also known as Kohonen's maps \cite{kohonen1990self} \cite{kohonen1998self}, are a special class of NNs that uses the concept of \textit{competitive learning}, in which output neurons compete amongst themselves to be activated in a real-valued output, results having only single neuron (or group of neurons), called \textit{winning neuron}. This is achieved by creating lateral inhibition connections (negative feedback paths) between neurons \cite{rosenblatt1958perceptron}. In this orientation, the network determines the winning neuron within several iterations; subsequently it is forced to reorganize itself based on the input data distribution (hence they are called Self-Organizing Maps). They were initially inspired by the human brain, which has specialized regions in which different sensory inputs are represented/processed by topologically ordered computational maps. In SOM, neurons are arranged on vertices of a lattice (commonly one or two dimensions). The network is forced to represent higher-dimensional data in lower-dimensional representation by preserving the topological properties of input data by using neighborhood function while transforming the input into a topological space in which neuron positions in the space are representatives of intrinsic statistical features that tell us about the inherent nonlinear nature of SOMs.
                            
Training a network comprising SOM is essentially a three-stage process after random initialization of weighted connections. The three stages are as follow \cite{haykin2009neural}.
                            
\begin{itemize}
\item \textit{Competition:} Each neuron in the network computes its value using a discriminant function, which provides the basis of competition among the neurons. Neuron with the largest discriminant value in the competition group is declared the winner.
                            
\item \textit{Cooperation:} The winner neuron then locates the center of the topological neighborhood of excited neurons in the previous stage, providing a basis for cooperation among excited neighboring neurons.
                            
\item \textit{Adaption:} The excited neurons in the neighborhood increase/decrease their individual values of discriminant function in regard to input data distribution through subtle adjustments such that the response of the winning neuron is enhanced for similar subsequent input. Adaption stage is distinguishable into two sub-stages: (1) the \textit{ordering or self-organizing phase}, in which weight vectors are reordered according to topological space; and (2) the \textit{convergence phase}, in which the map is fine-tuned and declared accurate to provide statistical quantification of the input space. This is the phase in which the map is declared to be converged and hence trained.
\end{itemize}
                        
One essential requirement in training a SOM is the redundancy of the input data to learn about the underlying structure of neuron activation patterns. Moreover, sufficient quantity of data is required for creating distinguishable clusters; withstanding enough data for classification problem, there exist a problem of gray area between clusters and creation of infinitely small clusters where input data has minimal patterns.

\vspace{2mm}
\textit{Adaptive Resonance Theory}: Adaptive Resonance Theory (ART) is another different category of NN models that is based on the theory of human cognitive information processing. It can be explained as an algorithm of incremental clustering which aims at forming multi-dimensional clusters, automatically discriminating and creating new categories based on input data. Primarily, ART models are classified as unsupervised learning model; however, there exist ART variants that employ supervised and hybrid learning approaches as well. The main setback of most NN models is that they lose old information (updating/diminishing weights) as new information arrives, therefore an ideal model should be flexible enough to accommodate new information without losing the old one, and this is called the \textit{plasticity-stability} problem. ART models provide a solution to this problem by self-organizing in real time and creating a competitive environment for neurons, automatically discriminating/creating new clusters among neurons to accommodate any new information.
                        
ART model resonates around (top-down) observer expectations and (bottom-up) sensory information while keeping their difference within the threshold limits of vigilance parameter, which in result is considered as the member of the expected class of neurons \cite{carpenter2010adaptive}. Learning of an ART model primarily consists of a comparison field, recognition field, vigilance (threshold) parameter and a reset module. The comparison field takes an input vector, which in result is passed, to best match in the recognition field; the best match is the current winning neuron. Each neuron in the recognition field passes a negative output in proportion to the quality of the match, which inhibits other outputs therefore exhibiting lateral inhibitions (competitions). Once the winning neuron is selected after a competition with the best match to the input vector, the reset module compares the quality of the match to the vigilance threshold. If the winning neuron is within the threshold, it is selected as the output, else the winning neuron is reset and the process is started again to find the next best match to the input vector. In case where no neuron is capable to pass the threshold test, a search procedure begins in which the reset module disables recognition neurons one at a time to find a correct match whose weight can be adjusted to accommodate the new match, therefore ART models are called self-organizing and can deal with the plasticity/stability dilemma.

\vspace{2mm}
\subsubsection{Unsupervised Deep NN}
	            
In recent years unsupervised deep NN has become the most successful unsupervised structure due to its application in many benchmarking problems and applications \cite{karhunen2015unsupervised}. Three major types of unsupervised deep NNs are deep belief NNs, deep autoencoders, and convolutional NNs. 
	            

\vspace{2mm}
\textit{Deep Belief NN}: Deep Belief Neural Network or simply Deep Belief Networks (DBN) is a probability based generative graph model that is composed of hierarchical layers of stochastic latent variables having binary valued activations, which are referred as hidden units or feature detectors. The top layers in DBNs have undirected, symmetric connections between them forming associative memory. DBNs provide a breakthrough in unsupervised learning paradigm. In the learning stage, DBN learns to reconstruct its input, each layer acting as feature detectors. DBN can be trained by greedy layer-wise training starting from the top layer with raw input, subsequent layers are trained with the input data from the previous visible layer \cite{hinton2006fast}. Once the network is trained in unsupervised manner and learned the distribution of the data, it can be fine tuned using supervised learning methods, or supervised layers can be concatenated in order to achieve the desired task (for instance, classification).
                

\vspace{2mm}
\textit{Deep Autoencoder}: Another famous type of DBN is the \textit{deep autoencoder}, which is composed of two symmetric DBNs---the first of which is used to encode the input vector, while the second decodes. By the end of the training of the deep autoencoder, it tends to reconstruct the input vector at the output neurons, and therefore the central layer between both DBNs is the actual compressed feature vector.
                

\vspace{2mm}
\textit{Convolutional NN}: Convolutional NN (CNN) are feed forward NN in which neurons are adapted to respond to overlapping regions in two-dimensional input fields such as visual or audio input. It is commonly achieved by local sparse connections among successive layers and tied shared weights followed by rectifying and pooling layers which results in transformation invariant feature extraction. Another advantage of CNN over simple multilayer NN is that it is comparatively easier to train due to sparsely connected layers with the same number of hidden units. CNNs represent the most significant type of architecture for computer vision as they solve two challenges with the conventional NNs: 1) scalable and computationally tractable algorithms are needed for processing high-dimensional images; and 2) algorithms should be transformation invariant since objects in an image can occur at an arbitrary position. However, most CNNs are composed of supervised feature detectors in the lower and middle hidden layers. In order to extract features in an unsupervised manner, a hybrid of CNN and DBN, called Convolutional Deep Belief Network (CDBN), is proposed in \cite{lee2009convolutional}. Making probabilistic max-pooling\footnote{Max-pooling is an algorithm of selecting the most responsive receptive field of a given interest region.} to cover larger input area and convolution as an inference algorithm makes this model scalable with higher dimensional input. Learning is processed in an unsupervised manner as proposed in \cite{bengio2007greedy}, i.e., greedy layer-wise (lower to higher) training with unlabeled data.

CDBN is a promising scalable generative model for learning translation invariant hierarchical representation from any high-dimensional unlabeled data in an unsupervised manner taking advantage of both worlds, i.e., DBN and CNN. CNN, being widely employed for computer vision applications, can be employed in computer networks for optimization of Quality of Experience (QoE) and Quality of Service (QoS) of multimedia content delivery over networks, which is an open research problem for next generation computer networks \cite{barakovic2013survey}.
	           
	    \begin{figure*}
        	\begin{center}
        		\includegraphics[width=.7\textwidth]{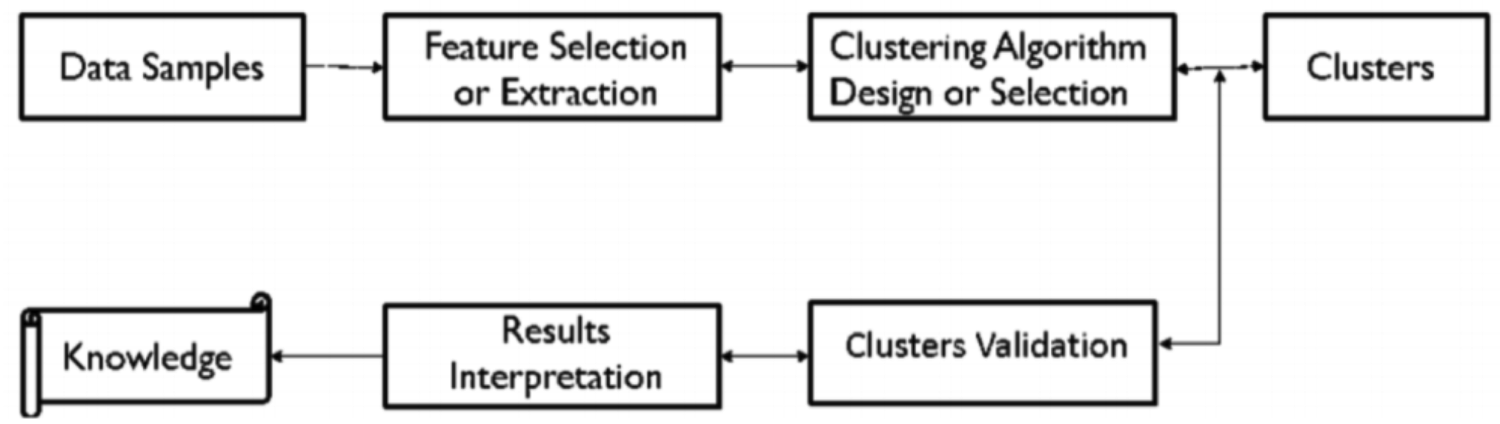}
        		\caption{Clustering process}
        		\label{fig:Clustering_eg}
        	\end{center}
        \end{figure*}
	    
\vspace{2mm}            
\subsubsection{Unsupervised Recurrent NN} \label{sec:cyclicNN}

Recurrent NN (RNN) is the most complex type of NN, and hence the nearest match to an actual human brain that processes sequential inputs. It can learn temporal behaviors of a given training data. RNN employs an internal memory per neuron to process such sequential inputs in order to exhibit the effect of previous event on the next. Compared to feed forward NNs, RNN is a stateful network. It may contain computational cycles among states, and uses time as the parameter in the transition function from one unit to another. Being complex and recently developed, it is an open research problem to create domain-specific RNN models and train them with a sequential data. Specifically, there are two perspectives of RNN to be discussed in the scope of this survey, namely, the depth of the architecture and the training of the network. The depth, in the case of a simple artificial NN, is the presence of hierarchical nonlinear intermediate layers between the input and output signals. In the case of a RNN, there are different hypotheses explaining the concept of depth. One hypothesis suggests that RNNs are inherently deep in nature when expanded with respect to sequential input; there are a series of nonlinear computations between the input at time $t(i)$ and the output at time $t(i+k)$.
                
However, at an individual discrete time step, certain transitions are neither deep nor nonlinear. There exist input-to-hidden, hidden-to-hidden, and hidden-to-output transitions, which are shallow in the sense that there are no intermediate nonlinear layers at discrete time step. In this regard, different deep architectures are proposed in \cite{pascanu2013construct} that introduce intermediate nonlinear transitional layers in between the input, hidden and output layers. Another novel approach is also proposed by stacking hidden units to create hierarchical representation of hidden units, which mimic the deep nature of standard deep NNs.
                
Due to the inherent complex nature of RNN, to the best of our knowledge, there is no widely adopted approach for training RNNs and many novel methods (both supervised and unsupervised) are introduced to train RNNs. Considering unsupervised learning of RNN in the scope of this paper, Klapper-Rybicka et al. \cite{klapper2001unsupervised} employ Long Short-term Memory (LSTM) RNN to be trained in an unsupervised manner using unsupervised learning algorithms, namely Binary Information Gain Optimization and Non-Parametric Entropy Optimization, in order to make a network to discriminate between a set of temporal sequences and cluster them into groups. Results have shown remarkable ability of RNNs for learning temporal sequences and clustering them based on a variety of features. Two major types of unsupervised recurrent NN are Hopfield NN and Boltzmann machine.
                

\vspace{2mm}
\textit{Hopfield NN}: Hopfield NN is a cyclic recurrent NN where each node is connected to other. Hopfield NN provides an abstraction of circular shift register memory with nonlinear activation functions to form a global energy function with guaranteed convergence to local minima. Hopfield NNs are used for finding clusters in the data without a supervisor.

\vspace{2mm}            
\textit{Boltzmann Machine}: Boltzmann machine is a stochastic symmetric recurrent NN that is used for search and learning problems. Due to binary vector based simple learning algorithm of Boltzmann machine, very interesting features representing the complex unstructured data can be learned \cite{Hinton:2007}. Since Boltzmann machine uses multiple hidden layers as feature detectors, the learning algorithm becomes very slow. To avoid the slow learning and to achieve faster feature detection instead of Boltzmann machine, a faster version, namely restricted Boltzmann machine (RBM), is used for practical problems \cite{salakhutdinov2009deep}. Restricted Boltzmann machine learns a probability distribution over its input data. It is faster than a Boltzmann machine because it only uses one hidden layer as feature detector layer. RBM is used for dimensionality reduction, clustering and feature learning in computer networks.

\vspace{2mm}	    
\subsubsection{Significant Applications of Hierarchical Learning in Networks}

ANNs/DNNs are the most researched topic when creating intelligent systems in computer vision and natural language processing whereas their application in computer networks are very limited, they are employed in different networking applications such as classification of traffic, anomaly/intrusion detection, detecting Distributed Denial of Service (DDoS) attacks, and resource management in cognitive radios \cite{tsagkaris2008neural}. The  motivation of using DNN for learning and predicting in networks is the unsupervised training that detects hidden patterns in ample amount of data that is near to impossible for a human to handcraft features catering for all scenarios. Moreover, many new research shows that a single model is not enough for the need of some applications, so developing a hybrid NN architecture having pros and cons of different models creates a new efficient NN which provides even better results. Such an approach is used in \cite{teles2007neural}, in which a hybrid model of ART and RNN is employed to learn and predict traffic volume in a computer network in real time. Real-time prediction is essential to adaptive flow control, which is achieved by using hybrid techniques so that ART can learn new input patterns without re-training the entire network and can predict accurately in the time series of RNN. Furthermore, DNNs are also being used in resource allocation and QoE/QoS optimizations. Using NN for optimization, efficient resource allocation without affecting the user experience can be crucial in the time when resources are scarce. Authors of \cite{munaretto2015data}, \cite{badia2014cognition} propose a simple DBN for optimizing multimedia content delivery over wireless networks by keeping QoE optimal for end users. Table \ref{HLDL} also provides a tabulated description of hierarchical learning in networking applications. However, these are just a few notable examples of deep learning and neural networks in networks, refer to Section \ref{sec:wwc} for more applications and detailed discussion on deep learning and neural networks in computer networks.

	     \begin{figure*}[ht]
        	\begin{center}
        		\includegraphics[width=.8\textwidth]{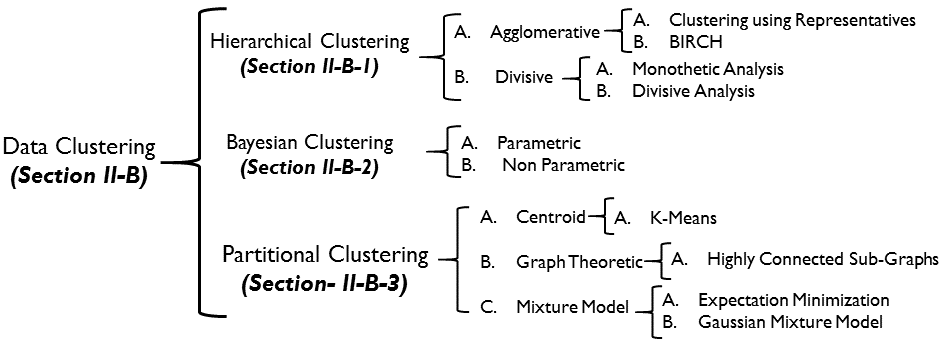}
        		\caption{Clustering Taxonomy}
        		
        		\label{fig:Clustering_overview}
        	
        	\end{center}
        \end{figure*}
        
\subsection{Data Clustering} 
\label{subsec:dataClustering}

Clustering is an unsupervised learning task that aims to find hidden patterns in unlabeled input data in the form of clusters \cite{grira2004unsupervised}. Simply put, it encompasses arrangement of data in meaningful natural groupings on the basis of the similarity between different \textit{features} (as illustrated in Figure \ref{fig:Clustering_eg}) to learn about its structure. Clustering involves the organization of data in such a way that there is high intra-cluster and low inter-cluster similarity. The resulting structured data is termed as \textit{data-concept} \cite{berkhin2006survey}. Clustering is used in numerous applications from the fields of ML, data mining, network analysis, pattern recognition and computer vision. The various techniques used for data clustering are described in more detail later in Section \ref{subsec:dataClustering}. In networking, clustering techniques are widely deployed for applications such as traffic analysis and anomaly detection in all kinds of networks (e.g., wireless sensor networks and mobile adhoc networks), with anomaly detection \cite{bhuyan2014network}. 

Clustering improves performance in various applications. McGregor et al. \cite{mcgregor2004flow} propose an efficient packet tracing approach using the Expectation-Maximization (EM) probabilistic clustering algorithm, which groups flows (packets) into a small number of clusters, where the goal is to analyze network traffic using a set of representative clusters.
        
A brief overview of different types of clustering methods and their relationships can be seen in Figure \ref{fig:Clustering_overview}. Clustering can be divided into three main types \cite{xu2005survey}, namely \textit{hierarchical clustering}, \textit{Bayesian clustering}, and \textit{partitional clustering}. Hierarchical clustering creates a hierarchical decomposition of data, whereas Bayesian clustering forms a probabilistic model of the data that decides the fate of a new test point probabilistically. In contrast, partitional clustering constructs multiple partitions and evaluates them on the basis of certain criterion or characteristic such as the Euclidean distance.  
                
Before delving into the general sub-types of clustering, there are two unique clustering techniques, which need to be discussed, namely \textit{density-based clustering} and \textit{grid-based clustering}. In some cases, density-based clustering is classified as a partitional clustering technique; however, we have kept it separate considering its applications in networking. Density-based models target the most densely populated area of a data space, and separates it from areas having low densities, thus forming clusters \cite{rehman2005comparison}. Chen and Tu \cite{chen2007density} use density-based clustering to cluster data stream in real time, which is important in many applications (e.g., intrusion detection in networks). Another technique is grid-based clustering, which divides the data space into cells to form a grid-like structure; subsequently, all clustering actions are performed on this grid \cite{leung2005unsupervised}. Leung and Leckie \cite{leung2005unsupervised} also present a novel approach that uses customized grid-based clustering algorithm to detect anomalies in networks. 
        
We move on next to describe three major types of data clustering approaches as per the taxonomy shown in Figure \ref{fig:Clustering_overview}.
       

\vspace{2mm}      
\subsubsection{Hierarchical Clustering}
	            
Hierarchical clustering is a well-known strategy in data mining and statistical analysis in which data is clustered into a hierarchy of clusters using an agglomerative (bottom up) or a divisive (top down) approach. Almost all hierarchical clustering algorithms are unsupervised and deterministic. The primary advantage of hierarchical clustering over unsupervised K-means and EM algorithms is that it does not require the number of clusters to be specified beforehand. However, this advantage comes at the cost of computational efficiency. Common hierarchical clustering algorithms have at least quadratic computational complexity compared to the linear complexity of K-means and EM algorithms. Hierarchical clustering methods have a pitfall: these methods fail to accurately classify messy high-dimensional data as its heuristic may fail due to the structural imperfections of empirical data. Furthermore, the computational complexity of the common agglomerative hierarchical algorithms is NP-hard. SOM, as discussed in Section \ref{competitiveNNs}, is a modern approach that can overcome the shortcomings of hierarchical models \cite{mangiameli1996comparison}.
        
\vspace{2mm}
\subsubsection{Bayesian Clustering} Bayesian clustering is a probabilistic clustering strategy where the posterior distribution of the data is learned on the basis of a prior probability distribution. Bayesian clustering is divided into two major categories, namely parametric and non-parametric \cite{orbanz2011bayesian}. Major difference between parametric and non-parametric techniques is the dimensionality of parameter space: if there are finite dimensions in the parameter space, the underlying technique is called Bayesian parametric; otherwise, the underlying technique is called Bayesian non-parametric. A major pitfall with the Bayesian clustering approach is that the choice of the wrong prior probability distributions can distort the projection of the data. Kurt et al. \cite{kurtbayesian} performed Bayesian nonparametric clustering of network traffic data to determine the network application type. 

\vspace{2mm}
\subsubsection{Partitional Clustering}
Partitional clustering corresponds to a special class of clustering algorithms that decomposes data into a set of disjoint clusters. Given $n$ observations, the clustering algorithm partitions a data into $k<n$ clusters \cite{Jin2010}. Partitional clustering is further classified into K-means clustering and mixture models.

\vspace{2mm}
\paragraph{K-Means Clustering}
K-means clustering is a simple, yet widely used approach for classification. It takes a statistical vector as an input to deduce classification models or classifiers. K-means clustering tends to distribute $m$ observations into $n$ clusters where each observation belongs to the nearest cluster. The membership of an observation to a cluster is determined using the cluster mean. K-means clustering is used in numerous applications in the domains of network analysis and traffic classification. Gaddam et al.  \cite{gaddam2007k} use K-means clustering in conjunction with supervised ID3 decision tree learning models to detect anomalies in a network. ID3 decision tree is an iterative supervised decision tree algorithm based on the concept learning system.  K-means clustering provided excellent results when used in traffic classification. Yingqiu et al. \cite{yingqiu2007network} show that K-means clustering performs well in traffic classification with an accuracy of 90\%. 
                
K-means clustering is also used in the domain of network security and intrusion detection. Meng et al. \cite{jianliang2009application} propose a K-means algorithm for intrusion detection. Experimental results on a subset of KDD-99 dataset show that detection rate stays above 96\% while the false alarm rate stays below 4\%. Results and analysis of experiments on K-means algorithm have demonstrated a better ability to search clusters globally.

Another variation of K-means is known as K-medoids, in which rather than taking the mean of the clusters, the most centrally located data point of a cluster is considered as the reference point of the corresponding  cluster \cite{chitrakar2012anomaly}. Few of the applications of K-medoids in the spectrum of anomaly detection can be seen here \cite{chitrakar2012anomaly} \cite{chitrakar2012anomalyhybrid}.
                
\vspace{2mm}    
\paragraph{Mixture Models}

Mixture models are powerful probabilistic models for univariate and multivariate data. Mixture models are used to make statistical inferences and deductions about the properties of the sub-populations given only observations on the pooled population. They are also used to statistically model data in the domains of pattern recognition, computer vision, ML, etc. Finite mixtures, which are a basic type of mixture model, naturally model observations that are produced by a set of alternative random sources. Inferring and deducing different parameters from these sources based on their respective observations lead to clustering of the set of observations. This approach to clustering tackles drawbacks of heuristic based clustering methods, and hence it is proven to be an efficient method for node classification in any large-scale network and has shown to yield efficient results compared to techniques commonly used. For instance, K-means and hierarchical agglomerative methods rely on supervised design decisions, such as the number of clusters or validity of models \cite{figueiredo2002unsupervised}. Moreover, combining EM algorithm with mixture models produces remarkable results in deciphering the structure and topology of the vertices connected through a multi-dimensional network \cite{newman2007mixture}. Bahrololum et al. \cite{bahrololum2008anomaly} used Gaussian mixture model (GMM) to outperform signature based anomaly detection in network traffic data. 
                    
\vspace{2mm}	            
\subsubsection{Significant Applications of Clustering in Networks}

Clustering can be found in mostly all unsupervised learning problems, and there are diverse applications of clustering in the domain of computer networks. Two major networking applications where significant use of clustering can be seen are intrusion detection and Internet traffic classification. One novel way to detect anomaly is proposed \cite{chimphlee2006anomaly}, and this approach preprocesses the data using Genetic Algorithm (GA) combined with hierarchical clustering approach called Balanced Iterative Reducing using Clustering Hierarchies (BIRCH) to provide an efficient classifier based on Support Vector Machine (SVM). This hierarchical clustering approach stores abstracted data points instead of the whole dataset, thus giving more accurate and quick classification compared to all past methods, producing better results in detecting anomalies. Another approach \cite{leung2005unsupervised} discusses the use of grid-based and density-based clustering for anomaly and intrusion detection using unsupervised learning. Basically, a scalable parallel framework for clustering large datasets with high dimensions is proposed and then improved by inculcating frequency pattern trees. Table \ref{DCNA} also provides a tabulated description of data clustering applications in networks. These are just few notable examples of clustering approaches in networks: refer to Section \ref{sec:wwc} for detailed discussion on some salient clustering applications in the context of networks.

 \begin{table*}
	\caption{Applications of Data Clustering in Networking Applications \label{DCNA}}
    {
	\centering
		\scriptsize
	\begin{tabular}{  p{5cm} p{2cm}  p{10cm}  }
		
		\toprule
		Reference &  Technique & Brief Summary\\
		\midrule
		
		\textit{\textbf{\underline{Internet Traffic Classification}}} \\
		\\
	    
	    Adda et al.\cite{adda2017comparative}&K-means \& EM&A comparative analysis of Network traffic fault classification is performed between K-means and EM techniques. \\
	    
	    Vluaductu et al.\cite{vluaductu2017internet}&K-means \& Dissimilarity-based clustering&Semi supervised approach for Internet traffic classification benefits from K-means and dissimilarity-based clustering as a first step for the Internet traffic classification. \\
	    
	    Liu et al.\cite{liu2017effective}&K-means&A novel variant of K-means clustering namely recursive time continuity constrained K-Means clustering, is proposed and used for real-time In-App activity analysis of encrypted traffic streams. Extracted feature vector of cluster centers are fed to random forest for further classification. \\

		\midrule
		\textit{\textbf{\underline{Anomaly/Intrusion Detection}}} \\
		\\
	    Parwez et al.\cite{parwez2017big}&K-means \& Hierarchical Clustering&K-means and hierarchical clustering is used to detect anomalies in call detail records of mobile wireless networks data.\\
	    
	    Lorido et al.\cite{lorido2017unsupervised}&GMM&GMM is used for detecting the anomalies that are affecting resources in cloud data centers. \\
	    
	    Frishman et al.\cite{frishman2017cluster}&K-means&K-means clustering is used for clustering the input data traffic for load balancing for network security. \\
	

		\midrule
	\textit{\textbf{\underline{Dimensionality Reduction and Visualization}}} \\
	\\
	
	Kumar et al.\cite{kumar2017feature}&Fuzzy Feature Clustering&A new feature clustering based approach for  dimensionality reduction of Internet traffic for intrusion detection is presented.\\
	
	Wiradinata et al.\cite{wiradinata2016clustering}&Fuzzy C-mean clustering \& PCA& This works combines data clustering technique combined with PCA is used for dimensionality reduction and classification of the Internet traffic. \\
		
	

		   \hline
		\end{tabular}
	}
\end{table*}   
                
\subsection{Latent Variable Models} 

A latent variable model is a statistical model that relates the manifest variables with a set of latent or hidden variables. Latent variable model allows us to express relatively complex distributions in terms of tractable joint distributions over an expanded variable space \cite{bishop1998latent}. Underlying variables of a process are represented in higher dimensional space using a fixed transformation, and stochastic variations are known as latent variable models where the distribution in higher dimension is due to small number of hidden variables acting in a combination \cite{skrondal2007latent}. These models are used for data visualization, dimensionality reduction, optimization, distribution learning, blind signal separation and factor analysis. Next we will begin our discussion on various latent variable models, namely \textit{mixture distribution}, \textit{factor analysis}, \textit{blind signal separation}, \textit{non-negative matrix factorization}, \textit{Bayesian networks \& probabilistic graph models (PGM)}, \textit{hidden Markov model (HMM)}, and \textit{nonlinear dimensionality reduction techniques} (which further includes \textit{generative topographic mapping}, \textit{multi-dimensional scaling}, \textit{principal curves}, \textit{Isomap}, \textit{localliy linear embedding}, and \textit{t-distributed stochastic neighbor embedding}).  

\vspace{2mm}
\subsubsection{Mixture Distribution}
Mixture distribution is an important latent variable model that is used for estimating the underlying density function. Mixture distribution provides a general framework for density estimation by using the simpler parametric distributions. Expectation maximization (EM) algorithm is used for estimating the mixture distribution model \cite{bishop1995neural}, through a maximization of the log likelihood of the mixture distribution model.
	    
\vspace{2mm}
\subsubsection{Factor Analysis}
Another important type of latent variable model is factor analysis, which is a density estimation model. It has been used quite often in collaborative filtering and dimensionality reduction. It is different from other latent variable models in terms of the allowed variance for different dimensions as most latent variable models for dimensionality reduction in conventional settings use a fixed variance Gaussian noise model.  In factor analysis model, latent variables have diagonal covariance rather than isotropic covariance. 

\vspace{2mm}
\subsubsection{Blind Signal Separation}
\begin{figure*}
            	\begin{center}
            		\includegraphics[width=.7\textwidth]{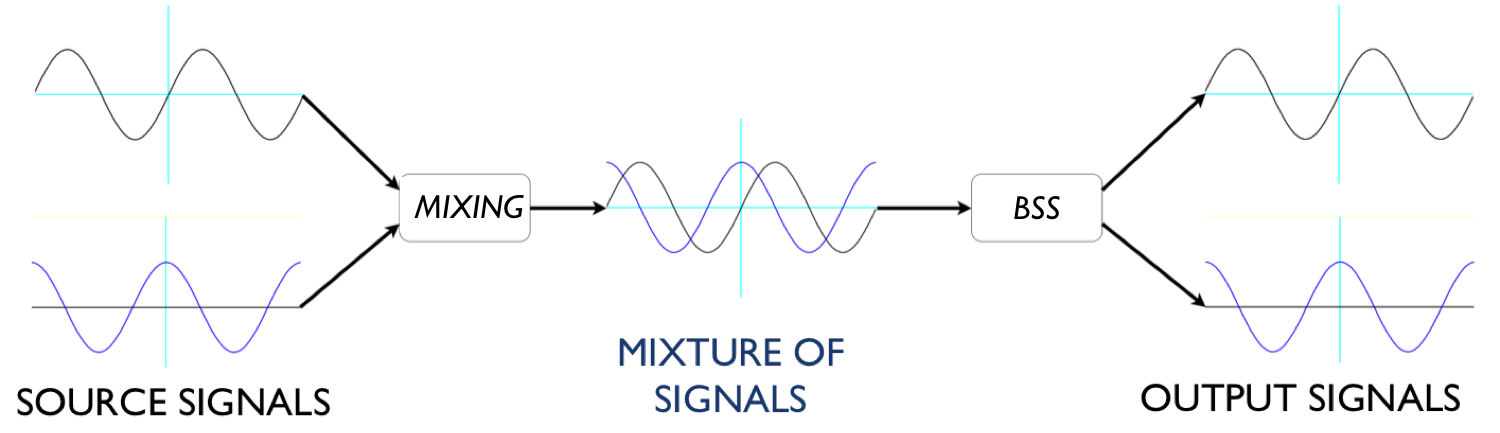}
            		\caption{Blind signal separation (BSS): A mixed signal composed of various input signals mixed by some mixing process is blindly processed (i.e., with no or minimal information about the mixing process) to show the original signals.}
            		\label{fig:bss}
            	\end{center}
            \end{figure*}

Blind Signal Separation (BSS), also referred to as Blind Source Separation, is the identification and separation of independent source signals from mixed input signals without or very little information about the mixing process. Figure \ref{fig:bss} depicts the basic BSS process in which source signals are extracted from a mixture of signals. It is a fundamental and challenging problem in the domain of signal processing although the concept is extensively used in all types of multi-dimensional data processing. Most common techniques employed for BSS are principal component analysis (PCA) and independent component analysis (ICA).
	        
\vspace{1mm}	       
\textit{a)} \textit{Principal Component Analysis} (PCA) is a statistical procedure that utilizes orthogonal transformation on the data to convert $n$ number of possibly correlated variables into lesser $k$ number of uncorrelated variables named principal components. Principal components are arranged in the descending order of their variability, first one catering for the most variable and the last one for the least. Being a primary technique for exploratory data analysis, PCA takes a cloud of data in $n$ dimensions and rotates it such that maximum variability in the data is visible. Using this technique, it brings out the strong patterns in the dataset so that these patterns are more recognizable thereby making the data easier to explore and visualize.
                
PCA has primarily been used for dimensionality reduction in which input data of $n$ dimensions is reduced to $k$ dimensions without losing critical information in the data. The choice of the number of principal components is a question of design decision. Much research has been conducted on selecting the number of components such as cross-validation approximations \cite{josse2012selecting}. Optimally, $k$ is chosen such that the ratio of the average squared projection error to the total variation in the data is less than or equal to 1\% by which 99\% of variance is retained in the $k$ principal components. But, depending on the application domain, different designs can increase/decrease the ratio while maximizing the required output. Commonly, many features of a dataset are often highly correlated; hence, PCA results in retaining 99\% of the variance while significantly reducing the data dimensions.

\vspace{1mm}	       	        
\textit{b)} \textit{Independent Component Analysis} (ICA) is another technique for BSS that focuses in separating multivariate input data into additive components with the underlying assumption that the components are non-Gaussian and statistically independent. The most common example to understand ICA is the \textit{cocktail party problem} in which there are $n$ people talking simultaneously in a room and one tries to listen to a single voice. ICA actually separates source signals from input mixed signal by either minimizing the statistical dependence or maximizing the non-Gaussian property among the components in the input signals by keeping the underlying assumptions valid. Statistically, ICA can be seen as the extension of PCA, while PCA tries to maximize the second moment (variance) of data, hence relying heavily on Gaussian features; on the other hand, ICA exploits inherently non-Gaussian features of the data and tries to maximize the fourth moment of linear combination of inputs to extract non-normal source components in the data \cite{hyvarinen2000independent}.

\vspace{2mm}
\subsubsection{Non-Negative Matrix Factorization}

Non-Negative Matrix Factorization (NMF) is a technique to factorize a large matrix into two or more smaller matrices with no negative values, that is when multiplied, it reconstructs the approximate original matrix. NMF is a novel method in decomposing multivariate data making it easy and straightforward for exploratory analysis. By NMF, hidden patterns and intrinsic features within the data can be identified by decomposing them into smaller chunks, enhancing the interpretability of data for analysis, with positivity constraints. However, there exist many classes of algorithms \cite{wang2013nonnegative} for NMF having different generalization properties, for example, two of them are analyzed in \cite{lee2001algorithms}, one of which minimizes the least square error and while the other focuses on the Kullback-Leibler divergence keeping algorithm convergence intact.

\vspace{2mm}             
\subsubsection{Hidden Markov Model}

Hidden Markov Model (HMM) are stochastic models of great utility, especially in domains where we wish to analyze temporal or dynamic processes such as speech recognition, primary users (PU) arrival pattern in cognitive radio networks (CRNs), etc. HMMs are highly relevant to CRNs since many environmental parameters in CRNs are not directly observable. An HMM-based approach can analytically model a Markovian stochastic process in which we do not access to the actual states, which are assumed to be unobserved or hidden; instead, we can observe a state that is stochastically dependent on the hidden state. It is for this reason that an HMM is defined to be a doubly stochastic process: first, the underlying stochastic process is not observable; and second, another stochastic process, dependent on the underlying stochastic process, that produces a sequence of observed symbols \cite{cappe2009inference}.

\vspace{2mm}            
\subsubsection{Bayesian Networks \& Probabilistic Graph Models (PGM)}

In Bayesian learning we try to find the posterior probability distributions for all parameter settings, in this setup, we ensure that we have a posterior probability for every possible parameter setting. It is computationally expensive but we can use complicated models with small dataset and still avoid overfitting. Posterior probabilities are calculated by dividing the product of sampling distribution and prior distribution by marginal likelihood; in simple words posterior probabilities are calculated using Bayes theorem. Basis of reinforcement learning was also derived by using Bayes theorem \cite{duff2002optimal}. Since Bayesian learning is computationally expensive a new research trend is approximate Bayesian learning \cite{beal2003variational}. Authors in \cite{minka2001family} has given a comprehensive survey of different approximate Bayesian inference algorithms. With the emergence of Bayesian deep learning framework the deployment of Bayes learning based solution is increasing rapidly.
             
Probabilistic graph modeling is a concept associated with Bayesian learning. A model representing the probabilistic relationship between random variables through a graph is known as probabilistic graph model (PGM). Nodes and edges in the graph represent a random variable and their probabilistic dependence, respectively. PGM are of two types: directed PGM and undirected PGM. Bayes networks also fall in the regime of directed PGM. PGM are used in many important areas such as computer vision, speech processing and communication systems. Bayesian learning combined with PGM and latent variable models forms a probabilistic framework where deep learning is used as a substrate for making improved learning architecture for recommender systems, topic modeling, and control systems \cite{wang2016towards}.

\vspace{2mm}
\subsubsection{Significant Applications of Latent Variable Models in Networks}

In \cite{dubois2011latent}, authors have applied latent structure on email corpus to find interpretable latent structure as well as evaluating its predictive accuracy on missing data task. A dynamic latent model for social network is represented in \cite{foulds2011dynamic}. A characterization of the end-to-end delay using a Weibull mixture model is discussed in \cite{hernandez2006weibull}. Mixture models for end host traffic analysis has been explored in \cite{agosta2013mixture}. BSS is a set of statistical algorithms that are widely used in different application domains to perform different tasks such as dimensionality reduction, correlating and mapping features, etc. Yan et al. \cite{yan2014principal} employ PCA for Internet traffic classification in order to separate different types of flows in a network packet stream. Similarly, authors of \cite{xu2005adaptive} employ PCA for feature learning and a supervised SVM classifier for classification in order to detect intrusion in an autonomous network system. Another approach for detecting anomalies and intrusions proposed in \cite{guan2009fast} uses NMF to factorize different flow features and cluster them accordingly. Furthermore, ICA has been widely used in telecommunication networks to separate mixed and noisy source signals for efficient service. For example, \cite{albataineh2013new} extends a variant of ICA called Efficient Fast ICA (EF-ICA) for detecting and estimating the symbol signals from the mixed CDMA signals received from the source endpoint.

In other literature, PCA uses a probabilistic approach to find the degree of confidence in detecting anomaly in wireless networks \cite{ahmed2005probabilistic}. Furthermore, PCA is also chosen as a method of clustering and designing Wireless Sensor Networks (WSNs) with multiple sink nodes \cite{chatzigiannakis2006hierarchical}. However, these are just a few notable examples of BSS in networks, refer to Section \ref{sec:wwc} for more applications and detailed discussion on BSS techniques in the networking domain.

Bayesian learning has been applied for classifying the Internet traffic, where Internet traffic is classified based on the posterior probability distributions. Real discretized conditional probability is used to construct a Bayesian classifier for early traffic identification in campus network has been proposed in \cite{gu2010early}. Host level intrusion detection using Bayesian networks is proposed in \cite{xu2008continuous}. Authors in  \cite{al2012feature} purposed a Bayesian learning based feature vector selection for anomalies classification in BGP. Port scan attacks prevention scheme using Bayesian learning approach is discussed in \cite{liu2008network}. Internet threat detection estimation system is presented in \cite{ishiguro2004internet}. A new approach towards outlier detection using Bayesian belief networks is described in \cite{janakiram2006outlier}. Application of Bayesian networks in MIMO systems has been explored in \cite{haykin2004bayesian}. Location estimation using Bayesian network in LAN is discussed in \cite{ito2005bayesian}. Similarly Bayes theory and PGM are both used in Low Density Parity Check (LDPC) and Turbo codes, which are the fundamental components of information coding theory. Table \ref{LNA} also provides a tabulated description of latent variable models applications in networking. 

\begin{table*}
	\caption{Applications of Latent Variable Models in Networking Applications \label{LNA}}
    {
	\centering
		\scriptsize
	\begin{tabular}{  p{5cm} p{2cm}  p{10cm} }
		
		\toprule
		Reference &  Technique & Brief Summary\\
		\midrule
		\textit{\textbf{\underline{Internet Traffic Classification}}} \\
		\\
		
		Liu et al.\cite{liu2016improved}&Mixture Distribution&An improved EM algorithm is proposed which derives a better GMM and used for the Internet traffic classification. \\
		
		Shi et al.\cite{shi2017efficient}&PCA&PCA based feature selection approach is used for the Internet traffic classification. Where PCA is employed for feature selection and irrelevant feature removal. \\
		
		Troia et al.\cite{troia2017identification}&NMF& NMF based models are applied on the data streams to find the traffic patterns which frequently occurs in network for identification and classification of tidal traffic patterns in metro area mobile network traffic. \\
    
		\midrule
		\textit{\textbf{\underline{Anomaly/Intrusion Detection}}} \\
		\\
		
	   Nie et al.\cite{nie2017modeling}&Bayesian Networks&Bayesian networks are employed for anomaly and intrusion detection such as DDoS attacks in cloud computing networks. \\
	   
	   Bang et al.\cite{bang2017anomaly}&Hidden Semi-Markov Model&Hidden semi-Markov model is used to detect LTE signalling attack. \\
	   
		\midrule
	\textit{\textbf{\underline{Network Operations, Optimization and Analytics}}} \\
	\\
	
		Chen et al.\cite{chen2017scalable}&Bayesian Networks&Scale-able Bayesian network models are used for data flow monitoring and analysis. \\
		
		Mokhtar et al.\cite{mokhtar2017big}&HMM&HMM and statistical analytic techniques combined with semantic analysis are used to propose a network management tool.\\
		
		\midrule
	\textit{\textbf{\underline{Dimensionality Reduction and Visualization}}} \\
	\\
	Furno et al.\cite{furno2017joint}&PCA \& Factor Analysis & PCA and factor analysis are used for dimensionality reduction and latent correlation identification in mobile traffic demand data. \\
	
	Malli et al.\cite{malli2017new}&PCA&PCA is used for dimensionality reduction and orthogonal coordinates of the social media profiles in ranking the social media profiles. \\

	

		   \hline
		\end{tabular}
	}
\end{table*}   
\subsection{Dimensionality Reduction}

Representing data in fewer dimensions is another well-established task of unsupervised learning. Real world data often have high dimensions---in many datasets, these dimensions can run into thousands, even millions, of potentially correlated dimensions \cite{roweis2000nonlinear}. However, it is observed that the intrinsic dimensionality (governing parameters) of the data is less than the total number of dimensions. In order to find the essential pattern of the underlying data by extracting intrinsic dimensions, it is necessary that the real essence is not lost; e.g., it may be the case that a phenomenon is observable only in higher-dimensional data and is suppressed in lower dimensions, these phenomena are said to suffer from the curse of dimensionality \cite{keogh2011curse}. While \textit{dimensionality reduction} is sometimes used interchangeably with \textit{feature selection} \cite{pudil1998novel}\cite{yu2003feature}, a subtle difference exists between the two \cite{hartmann2006dimension}. Feature selection is traditionally performed as a supervised task with a domain expert helping in handcrafting a set of critical features of the data. Such an approach generally can perform well but is not scalable and prone to judgment bias. Dimensionality reduction, on the other hand, is more generally an unsupervised task, where instead of choosing a subset of features, it creates new features (dimensions) as a function of all features. Said differently, feature selection considers supervised data labels, while dimensionality reduction focuses on the data points and their distributions in N-dimensional space.
            
There exist different techniques for reducing data dimensions \cite{fodor2002survey} including projection of higher dimensional points onto lower dimensions, independent representation, and sparse representation, which should be capable of reconstructing the approximate data. Dimensionality reduction is useful for data modeling, compression, and visualization. By creating representative functional dimensions of the data and eliminating redundant ones, it becomes easier to visualize and form a learning model. Independent representation tries to disconnect the source of variation underlying the data distribution such that the dimensions of the representation are statistically independent \cite{goodfellow2016deep}. Sparse representation technique represents the data vectors in linear combinations of small basis vectors. 

It is worth noting here that many of the latent variable models (e.g., PCA, ICA, factor analysis) also function as techniques for dimensionality reduction. In addition to techniques such as PCA, ICA---which infer the latent inherent structure of the data through a linear projection of the data---a number of nonlinear dimensionality reduction techniques have also been developed and will be focused upon in this section to avoid repetition of linear dimensionality reduction techniques that have already been covered as part of the previous subsection. Linear dimensionality reduction techniques are useful in many settings but these methods may miss important nonlinear structure in the data due to their subspace assumption, which posits that the high-dimensional data points lie on a linear subspace (for example, on a 2-D or 3D plane). Such an assumption fails in high dimensions when data points are random but highly correlated with neighbors. In such environments nonlinear dimensionality reductions through \textit{manifold learning} techniques---which can be construed as an attempt to generalize linear frameworks like PCA so that nonlinear structure in data can also be recognized---become desirable. Even though some supervised variants also exist, manifold learning is mostly performed in an unsupervised fashion using the nonlinear manifold substructure learned from the high-dimensional structure of the data from the data itself without the use of any predetermined classifier or labeled data. Some nonlinear dimensionality reduction (manifold learning) techniques are described below:  

\subsubsection{Isomap}
Isomap is a nonlinear dimensionality reduction technique that finds the underlying low dimensional geometric information about the dataset. Algorithmic features of PCA and MDS are combined to learn the low dimensional nonlinear manifold structure in the data \cite{tenenbaum2000global}. Isomap uses geodesic distance along the shortest path to calculate the low dimension representation shortest path, which can be computed using Dijkstra's algorithm. 

\subsubsection{Generative Topographic Model}
Generative topographic mapping (GTM) represents the nonlinear latent variable mapping from continuous low dimensional distributions embedded in high dimensional spaces \cite{gtm-the-generative-topographic-mapping}. Data space in GTM is represented as reference vectors and these vectors are a projection of latent points in data space. It is a probabilistic variant of SOM and works by calculating the Euclidean distance between data points. GTM optimizes the log likelihood function, and the resulting probability defines the density in data space.

\subsubsection{Locally Linear Embedding}
Locally linear embedding (LLE) \cite{roweis2000nonlinear} is an unsupervised nonlinear dimensionality reduction algorithm. LLE represents the data in lower dimensions yet preserving the higher dimensional embedding. LLE depicts data in single global coordinate of lower dimensional mapping of input data. LLE is used to visualize multi-dimensional dimensional manifolds and feature extraction. 

\subsubsection{Principal Curves}
Principal curves is a nonlinear dataset summarizing technique where non-parametric curves passes through the middle of multi-dimensional dataset providing the summary of the dataset \cite{hastie1989principal}. These smooth curves minimize the average squared orthogonal distance between data points, this process also resembles to the maximum likelihood for nonlinear regression in the presence of Gaussian noise \cite{leeestimations}.

\subsubsection{Nonlinear Multi-dimensional Scaling}
Nonlinear multi-dimensional scaling (NMDS) \cite{kruskal1964nonmetric} is a nonlinear latent variable representation scheme. It works as an alternative scheme for factor analysis. In factor analysis, a multivariate normal distribution is assumed and similarities between different objects are expressed as a correlation matrix. Whereas NMDS does not impose such a condition, and it is designed to reach the optimal low dimensional configuration where similarities and dissimilarities among matrices can be observed. NMDS is also used in data visualization and mining tools for depicting the multi-dimensional data in 3 dimensions based on the similarities in the distance matrix.   

\subsubsection{t-Distributed Stochastic Neighbor Embedding}
t-distributed stochastic neighbor embedding (t-SNE) is another nonlinear dimensionality reduction scheme. It is used to represent high dimensional data in 2 or 3 dimensions. t-SNE constructs a probability distribution in high dimensional space and constructs a similar distribution in lower dimensions and minimizes the Kullback–Leibler (KL) divergence
between two distributions (which is a useful way to measure the difference between two probability distributions) \cite{maaten2008visualizing}. 

Table \ref{DRNA} also provides a tabulated description of dimensionality reduction applications in networking. The applications of nonlinear dimensionality reduction methods are later described in detail in Section \ref{DRapplications}.

\begin{table*}
	\caption{Applications of Dimensionality Reduction in Networking Applications \label{DRNA}}
    {
	\centering
		\scriptsize
	\begin{tabular}{ p{5cm} p{2cm}  p{10cm} }
		
		\toprule
		Reference &  Technique & Brief Summary\\
		\midrule
		\textit{\textbf{\underline{Internet Traffic Classification}}} \\
		\\
		
	Cao et al.\cite{cao2017accurate}&PCA \& SVM& Internet traffic classification model is proposed based on PCA and SVM, where PCA is employed for dimensionality reduction and SVM for classification. \\
	
	Zhou et al.\cite{zhou2017som}&SOM \& Probabilistic NN&Proposed approach probabilistic neural network is used for dimensionality reduction and SOM are employed for network traffic classification.\\

		\midrule
		\textit{\textbf{\underline{Anomaly/Intrusion Detection}}} \\
		\\
	Erfani et al.\cite{erfani2016high}&DBN& Dimensionality reduction of high dimensional feature set is performed by training a DBN as nonlinear dimensionality reduction tool for human activity recognition using smart phones.\\
	
	Nicolau et al.\cite{nicolau2016hybrid}&Autoencoders&Latent representation learnt by using autoencoder is used for anomaly detection in network traffic, which is performed by using single Gaussian and full kernel density estimation. \\
	
	Ikram et al.\cite{ikram2016improving}&PCA \& SVM& A hybrid approach for intrusion detection is described, where PCA is used to perform dimensionality reduction operation on network data and SVM is used to detect intrusion in that low dimensional data.\\
	   
		\midrule
	\textit{\textbf{\underline{Network Operations, Optimization and Analytics}}} \\
	\\
	Moysen et al.\cite{moysen2017mobile}&PCA&PCA is used for low dimensional feature extraction in a mobile network planning tool based on data analytic.\\
	
	Ossia et al.\cite{ossia2017hybrid}&PCA \& Simple Embedding& PCA combined with simple embedding from deep learning is used for dimensionality reduction which reduces the communication overhead between client and server. \\
		
		\midrule
	\textit{\textbf{\underline{Dimensionality Reduction and Visualization}}} \\
	\\
	
	Rajendran et al.\cite{rajendran2017distributed}&t-SNE \& LSTM& LSTM is applied for modulation recognition in wireless data. t-SNE is used to perform dimensionality reduction and visualization of the wireless dataset's FFT response.\\
	
	Sarshar et al.\cite{sarshar2017analyzing}&t-SNE \& K-means& t-SNE is used for visualizing a high dimensional Wi-Fi mobility data in 3D. \\ 
		
	

		   \hline
		\end{tabular}
	}
\end{table*} 

\subsection{Outlier Detection} 

Outlier detection is an important application of unsupervised learning. A sample point that is distant from other samples is called an outlier. An outlier may occur due to noise, measurement error, heavy tail distributions and mixture of two distributions.  There are two popular underlying techniques for unsupervised outlier detection upon which many algorithms are designed, namely nearest neighbor based technique and clustering based method.

\vspace{2mm}
\subsubsection{Nearest Neighbor Based Outlier Detection}

Nearest neighbor method works on estimating the Euclidean distances or average distance of every sample from all other samples in the dataset. There are many algorithms based on nearest neighbor based techniques, with the most famous extension of nearest neighbor being k-nearest neighbor technique in which only k nearest neighbors participate in the outlier detection \cite{ramaswamy2000efficient}. Local outlier factor is another outlier detection algorithm, which works as an extension of the k-nearest neighbor algorithm. Connectivity based outlier factors \cite{tang2002enhancing}, influenced outlierness \cite{jin2006ranking}, and local outlier probability models \cite{kriegel2009loop} are few famous examples of the nearest neighbor based techniques.

\vspace{2mm}            
\subsubsection{Cluster Based Outlier Detection}

Clustering based methods use the conventional K-means clustering technique to find the dense locations in the data and  then perform density estimation on those clusters. After density estimation, a heuristic is used to classify the formed cluster according to the cluster size. Anomaly score is computed by calculating the distance between every point and its cluster head. Local density cluster based outlier factor \cite{he2003discovering}, clustering based multivariate Gaussian outlier score \cite{goldstein2014behavior}\cite{goldstein2016comparative} and histogram based outlier score \cite{goldstein2012histogram} are the famous cluster based outlier detection models in literature. SVM and PCA are also suggested for outlier detection in literature. 

\vspace{2mm}          
\subsubsection{Significant Applications of Outlier Detection in Networks}
Outlier detection algorithms are used in many different applications such as intrusion detection, fraud detection, data leakage prevention, surveillance, energy consumption anomalies, forensic analysis, critical state detection in designs, electrocardiogram and computed tomography scan for tumor detection. Unsupervised anomaly detection is performed by estimating the distances and densities of the provided non-annotated data \cite{chandola2009anomaly}. More applications of outlier detection schemes will be discussed in Section \ref{sec:wwc} 

\subsection{Reinforcement Learning}

Unsupervised learning can also be applied in the context of optimization and decision-making. Reinforcement Learning (RL) is an ML technique that attempts to learn about the optimal action with respect to the dynamic operating environment \cite{sutton1998RL}. Specifically, a decision maker (or an agent) observes state and reward from the operating environment and takes the best-known action, which leads to the optimal action as time goes by. Due to the dynamicity of the operating environment, the optimal action for the operating environment is expected to change; hence the need to learn about the optimal action from time to time. The state represents the decision-making factors, and the reward represents the positive or negative effects of the selected action on the network performance. For each state-action pair, an agent keeps track of its \textit{Q-value}, which accumulates the rewards for the action taken under the state, as time goes by. The agent selects an optimal action, which has the highest Q-value, in order to optimize the network performance. RL techniques can be broadly categorized as being either \emph{model-free} or \emph{model-based} \cite{sutton1998reinforcement}. We use the term model to refer to an abstraction used by the agent to predict how the environment will respond to its actions---i.e., given the state and the action performed therein by the agent, the model predicts stochastically the next state and the expected reward. 

To apply RL, the RL model (embedded in each agent) is identified by defining the state, action, and reward representations; this allows an agent access to a range of traditional and extended RL algorithms, such as the multi-agent approach. Most applications that apply RL take advantage of the benefits brought about by its intrinsic characteristics. Notably, RL takes account of a wide range of dynamic factors (e.g., traffic characteristics and channel capacity) affecting the network performance (e.g., throughput) since the reward represents the effects to the network performance. Also, RL does not need a model of the operating environment. This means that an agent can learn without prior knowledge about the operating environment. Nevertheless, the traditional RL approach comes with some shortcomings, particularly its inability to achieve network-wide performance enhancement, large number of state-action pairs, and low convergence rate to the optimal action. 

In recent times, there has been exciting developments in combining RL and deep neutral networks to create a more powerful hybrid approach called ``deep reinforcement learning'' that is also applicable to environments in which there are no handcrafted features available or where state spaces are not fully observed and low dimensional. Such techniques have been used to achieve human-level control that comfortably surpassed the performance of previous algorithms and achieved a level compared to professional human games tester across a set of 49 games including Atari 2600 games, using the same algorithm, architecture, and hyper-parameters \cite{mnih2015human}. The generality of such an approach can be used profitably and applied in the future in a number of networking settings. Next, we show some popular extended RL models that have been adopted to address the shortcomings of the traditional RL approach. 

\vspace{2mm}
\subsubsection{Multi-agent Reinforcement Learning}
While the traditional RL approach enables an individual agent to learn about the optimal action that maximizes the local network performance, Multi-agent Reinforcement Learning (MARL) enables a set of agents to learn about each other's information, such as Q-values and rewards, via direct communication or prediction to learn about the optimal joint action that maximizes the global performance \cite{schwartz2014multi}. A notable difference between MARL and the traditional RL approach is that both own and neighbors' information is used to update Q-values in MARL, while only own information is used in the traditional RL approach. By using the neighbor agents' information in the update of the Q-values, an agent takes account of the actions taken by its neighbor agents. This is necessary because an agent's action can affect and be affected by other agents' choice of actions in a shared operating environment. As time goes by, the agents select their respective action that is part of the joint action, which maximizes the global Q-value (or network-wide performance) in a collaborative manner. Various kinds of information can be exchanged including the Q-value of the current action \cite{Khan2012RL} and the maximum Q-value of the current state (also called value function) \cite{Schneider1999RL}. 

\vspace{2mm}
\subsubsection{Reinforcement Learning with Function Approximation}
The traditional RL approach keeps track of the Q-values of all state-action pairs in a tabular format. The number of state-action pairs grows exponentially as the number of states and actions grow, resulting in increased stress on the storage requirement of the Q-values. RL with function approximation (RLFA) represents the Q-values of the state-action pairs using a significantly smaller number of features. Each Q-value is represented using a feature, which consists of a set of measurable properties of a state-action pair, and a weight vector, which consists of a set of tunable parameters used to adjust the appropriateness of the feature \cite{Lunden2011RL}.     

\vspace{2mm}
\subsubsection{Model-based Reinforcement Learning}
During normal operation, an agent must converge to the optimal action; however, the convergence rate can be unpredictable due to the dynamicity of the operating environment. While increasing the learning rate (or the dependence on the current reward rather than historical rewards) of the RL model can intuitively increase the convergence rate, this can lead to the fluctuation of the Q-values if the current reward changes significantly particularly when the dynamicity of the operating environment is high. The model-based RL (MRL) approach addresses this by creating a model of the operating environment, and uses it to compute and update its Q-values. One way to do this is to estimate the state transition probability, which is the probability of a transition from one state to another when an action is undertaken \cite{Schneider1999RL}. Another way to do this is to compute the probability of the environment operating in a particular state \cite{Bkassiny2011RL}. The model of the operating environment can also serves as a learning tool. 

\vspace{2mm}
\subsubsection{Q-learning}
Q-learning, proposed by Watkins in 1992 \cite{watkins1992q}, is a popular \emph{model-free RL approach} that allows an agent to learn how to act optimally with comparatively little computational requirements. In a Q-learning setting, the agent directly determines the optimal policy by mapping environmental states to actions without constructing the corresponding stochastic model of the environment \cite{sutton1998reinforcement}. Q-learning works by incrementally improving its estimation of the \emph{Q-values}, which describe the quality of particular actions at particular states estimated by learning a \emph{Q-function} that gives the expected utility of taking a given action in a given state but following the optimal policy thereafter.

\vspace{2mm}            
\subsubsection{Significant Applications of RL in Networks} 
	
RL has been applied in wide ranges of applications to optimize network operations due to its versatility. Using MARL, agents exchange information (e.g., actions, Q-values, value functions) among themselves to perform target tracking where agents schedule and allocate target tracking tasks among themselves to keep track of moving objects in a WSN \cite{Khan2012RL}. Using RLFA, an agent reduces the large number of state-action pairs, which represent the probability of a channel being available and selected for transmission in channel sensing \cite{Lunden2011RL}. Using MRL, an agent can compute the state transition probability, which is used to select a next-hop node for packet transmission in routing \cite{Hu2010RLi}. Another application of MRL is to compute the probability of the operating environment operating in a particular state, which is then used to select a channel to sense and access in order to reduce interference. RL has also been proposed as an aid for enhancing security schemes for CRNs through the detection of malicious nodes and their attacks they launch \cite{ling2015application}. Q-learning is another popular RL technique that has been applied in the networking context---e.g., we highlight one example application of Q-learning in the context of Heterogeneous Mobile Networks (HetNets) \cite{pervez2017fuzzy} in which the authors proposed a fuzzy Q-learning based user-centric cell association scheme for ensuring appropriate QoS provisioning for users with results improving the state of the art.

\subsection{Lessons Learnt}
Key lessons drawn from the review of unsupervised learning techniques are summarized below. 
\begin{enumerate}
    
\item Hierarchical learning techniques are the most popular schemes in literature for feature detection and extraction. 

    
    
\item Learning the joint distribution of a complex distribution over an expanded variable space is a difficult task. Latent variable models have been the recommended and well-established schemes in literature for this problem. These models are also used for dimensionality reduction and better representation of data. 

\item Visualization of unlabeled multidimensional data is another unsupervised task. In this research we have explored the dimensionality reduction as a underlying scheme for developing a better multidimensional data visualization tools.
    
\item Reinforcement learning schemes for learning, decision-making, and network performance evaluation have also been surveyed and its potential application in network management and optimization is considered a potential research area. 

\end{enumerate}

\section{Applications of Unsupervised Learning in Networking} 
\label{sec:wwc}
	
In this section, we will introduce some significant applications of the unsupervised learning techniques that have been discussed in Section \ref{sec:technique} in the context of computer networks. We highlight the broad spectrum of applications in networking and emphasize the importance of ML-based techniques, rather than classical hard-coded statistical methods, for achieving more efficiency, adaptability, and performance enhancement.
	

           \begin{table*}[]	
           	\caption{Internet Traffic Classification with respect to Unsupervised Learning Techniques and Tasks \label{Table:Traffic Classification Systems}}
    {
           \centering
		\scriptsize
	\begin{tabular}{ p{3cm} p{2.5cm} p{2.5cm} p{8cm}  }
		\toprule
		Reference &  Technique  & Task & Brief Summary  \\
		\midrule
		\midrule
        
    
         Zhang et al. \cite{zhang2013network} &Non Parametric NN &Hierarchical Representations/ Deep Learning &Applied statistical correlation with non parametric NN to produce efficient and adaptive results in traffic classification.\\
         
		McGregor et al. \cite{mcgregor2004flow} &EM-based clustering &Data clustering &Applied EM probabilistic algorithm to cluster flows based on various attributes such as byte counts, inter-arrival statistics, etc. in flow classification.\\
		 
		 Erman et al. \cite{erman2006qrp05}&EM-based clustering &Data clustering &Applied EM-based clustering approach to yield 9\% better results compared to supervised Na{\"\i}ve Bayes based approach in traffic classification. \\
		 
		 Yingqiu et al. \cite{yingqiu2007network} &K-Means &Data clustering &Applied K-means clustering algorithm to produce an overall 90\% accuracy in Internet traffic classification in a completely unsupervised manner.\\
		 
		 Kornycky et al. \cite{kornycky2017radio} &GMM &Data Clustering &GMM with universal background model is used for encrypted WLAN traffic classification.\\
		 
		 Liu et al. \cite{liu2016real} &GMM &Data Clustering &GMM and Kerner's traffic theory based ML model is used to evaluate real-time Internet traffic performance. \\
		 
		 Erman et al. \cite{erman2006traffic} &K-Means, DBSCAN &Data clustering &Applied cluster analysis to effectively identify similar traffic using transport layer statistics to overcome the problem of dynamic port allocation in port based classification. \\
		 
		 Guyen et al. \cite{nguyen2008clustering} & Na{\"\i}ve Bayes clustering &Data clustering &Applied Na{\"\i}ve Bayes clustering algorithm in traffic classification.\\

		
		Yan et al. \cite{yan2014principal} &PCA &Blind Signal Separation &Applied PCA and fast correlation based filter algorithm that yields more accurate and stable experimental results in Internet traffic flow classification. 
		
		   \\
		   \hline

	\end{tabular}
	}
\end{table*}

\subsection{Internet Traffic Classification}
Internet traffic classification is of prime importance in networking as it provides a way to understand, develop and measure the Internet. Internet traffic classification is an important component for service providers to understand the characteristics of the service such as quality of service, quality of experience, user behavior, network security and many other key factors related to overall structure of the network \cite{shafiq2016network}. In this subsection, we will survey the unsupervised learning applications in network traffic classification.

As networks evolve at a rapid pace, the malicious intruders are also evolving their strategies. Numerous novel hacking and intrusion techniques are being regularly introduced causing severe financial jolts to companies and headaches to their administrators. Tackling these unknown intrusions through accurate traffic classification on the network edge therefore becomes a critical challenge and an important component of network security domain. Initially, when networks used to be small, simple port based classification technique that tried to identify the associated application with the corresponding packet based on its port number was used. However, this approach is now obsolete because recent malicious softwares use dynamic port-negotiation mechanism to bypass firewalls and security applications. A number of contrasting Internet traffic classification techniques have been proposed since then, and some important ones are discussed next.

Most of the modern traffic classification methods use different ML and clustering techniques to produce accurate clusters of packets depending on their applications, thus producing efficient packet classification \cite{nguyen2008survey}. The main purpose of classifying network's traffic is to recognize the destination application of the corresponding packet and to control the flow of the traffic when needed such as prioritizing one flow over others. Another important aspect of traffic classification is to detect intrusions and malicious attacks or screen out forbidden applications (packets).

First step in classifying Internet traffic is selecting accurate features, which is an extremely important, yet complex task. Accurate feature selection helps ML algorithms to avoid problems like class imbalance, low efficiency and low classification rate. There are three major feature selection methods in Internet traffic for classification: namely, the filter method, the wrapper based method, and the embedded method. These methods are based on different ML and genetic learning algorithms \cite{dhote2015survey}. Two major concerns in feature selection for Internet traffic classification are the large size of data and imbalanced traffic classes. To deal with these issues and to ensure accurate feature selection, a min-max ensemble feature selection scheme is proposed in \cite{huang2016internet}. A new information theoretic approach for feature selection for skewed datasets is described in \cite{zhen2012new}. This algorithm has resolved the multi-class imbalance issue but it does not resolve the issues of feature selection. In 2017, an unsupervised autoencoder based scheme has outperformed previous feature learning schemes, autoencoders were used as a generative model and were trained in a way that the bottleneck layer learnt a latent representation of the feature set; these features were then used for malware classification and anomaly detection to produce results that improved the state of the art in feature selection \cite{yousefi2017autoencoder}. 
        
Much work has been done on classifying traffic based on supervised ML techniques. Initially in 2004, the concept of clustering bi-directional flows of packets came out with the use of EM probabilistic clustering algorithm, which clusters the flows depending on various attributes such as packet size statistics, inter-arrival statistics, byte counts, and connection duration, etc. \cite{mcgregor2004flow}. Furthermore, clustering is combined with the above model \cite{nguyen2008clustering}; this strategy uses Na{\"\i}ve Bayes clustering to classify traffic in an automated fashion. Recently, unsupervised ML techniques have also been introduced in the domain of network security for classifying traffic. Major developments include a hybrid model to classify traffic in more unsupervised manner \cite{erman2007offline}, which uses both labeled and unlabeled data to train the classifier making it more durable and efficient. However, later on, completely unsupervised methods for traffic classification have been proposed, and still much work is going on in this area. Initially, completely unsupervised approach for traffic classification was employed using K-means clustering algorithm combined with log transformation to classify data into corresponding clusters. Then, \cite{yingqiu2007network} highlighted that using K-means and this method for traffic classification can improve accuracy by 10\% to achieve an overall 90\% accuracy.

Another improved and faster approach was proposed in 2006 \cite{bernaille2006traffic}, which examines the size of the first five packets and determines the application correctly using unsupervised learning techniques. This approach has shown to produce better results than the state-of-the-art traffic classifier, and also has removed its drawbacks (such as dealing with outliers or unknown packets, etc.). Another similar automated traffic classifier and application identifier can be seen in \cite{zander2005automated}, and they use the auto-class unsupervised Bayesian classifier, which automatically learns the inherent natural classes in a dataset.
        
In 2013, another novel strategy for traffic classification known as \textit{network traffic classification using correlation} was proposed \cite{zhang2013network}, which uses non-parametric NN combined with statistical measurement of correlation within data to efficiently classify traffic. The presented approach addressed the three major drawbacks of supervised and unsupervised learning classification models: \textit{firstly}, they are inappropriate for sparse complex networks as labeling of training data takes too much computation and time; \textit{secondly}, many supervised schemes such as SVM are not robust to training data size; and \textit{lastly}, and most importantly, all supervised and unsupervised algorithms perform poorly if there are few training samples. Thus, classifying the traffic using correlations appears to be more efficient and adapting. Oliveira et al. \cite{oliveira2016computer} compared four ANN approaches for computer network traffic, and modeled the Internet traffic as a time series and used mathematical methods to predict the time series. A greedy layer-wise training for  unsupervised stacked autoencoder produced excellent classification results, but at the cost of significant system complexity. Genetic algorithm combined with constraint clustering process are used for Internet traffic data characterization \cite{shrivastava2016internet}. In another work, a two-phased ML approach for Internet traffic classification using K-means and C5.0 decision tree is presented in \cite{bakhshi2016internet} where the average accuracy of classification was 92.37\%.
        
A new approach for Internet traffic classification has been introduced in 2017 by Vl{\u{a}}du{\c{t}}u et al. \cite{vluaductu2017internet} in which unidirectional and bidirectional information is extracted from the collected traffic, and K-means clustering is performed on the basis of statistical properties of the extracted flows. A supervised classifier then classifies these clusters. Another unsupervised learning based algorithm for Internet traffic detection is described in \cite{yang2017improved} where a restricted Boltzmann machine based SVM is proposed for traffic detection, this paper models the detection as classification problem. Results were compared with ANN and decision tree algorithms on the basis of precision and F1 score. Application of deep learning algorithms in Internet traffic classification has been discussed in \cite{fadlullah2017state}, with this work also outlining the open research challenges in applying deep learning for Internet traffic classification. These problems are related to training the models for big data since Internet data for deep learning falls in big data regime, optimization issues of the designed models given the uncertainty in Internet traffic and scalability of deep learning architectures in Internet traffic classification. To cope with the challenges of developing a flexible high-performance platform that can capture data from a high speed network operating at more than 60 Gbps, Gonzalez et al. \cite{gonzalez2017net2vec} have introduced a platform for high speed packet to tuple sequence conversion which can significantly advance the state of the art in real-time network traffic classification. In another work, Aminanto and Kim \cite{aminantodeep} used stacked autoencoders for Internet traffic classification and produced more than 90\% accurate results for the two classes in KDD 99 dataset.

Deep belief network combined with Gaussian model employed for Internet traffic prediction in wireless mesh backbone network has been shown to outperform the previous maximum likelihood estimation technique for traffic prediction \cite{nie2017network}. Given the uncertainty of WLAN channel traffic classification is very tricky, \cite{kornycky2017radio} proposed a new variant of Gaussian mixture model by incorporating universal background model and used it for the first time to classify the WLAN traffic. A brief overview of the different Internet traffic classification systems, classified on the basis of unsupervised technique and tasks discussed earlier, is presented in the Table \ref{Table:Traffic Classification Systems}.

\begin{table*}
	\caption{Anomaly \& Network Intrusion Detection Systems (A-NIDS) with respect to Unsupervised Learning Techniques \label{Table:ANIDS Systems}}
    {
	\centering
		\scriptsize
	\begin{tabular}{ p{2.7cm} p{3.3cm}  p{11cm}  }
		
		\toprule
		Reference &  Technique & Brief Summary\\
		\midrule
		
		\multicolumn{3}{l}{\textbf{\emph{\underline{Hierarchical Representations/ Deep Learning}}}}\\\\

	    Zhang et al. \cite{zhang2005intrusion}&Hierarchical NN &Applied radial basis function in a two layered hierarchical IDS to detect intruders in real time.\\	
		Rhodes et al. \cite{rhodes2000multiple} &SOM &Advocated unsupervised NNs such as SOM to provide a powerful supplement to existing IDSs. \\
		Kayacik et al. \cite{kayacik2003capability} &SOM &Overviewed the capabilities of SOM and its application in IDS. \\
		Zanero \& Stefano \cite{zanero2005analyzing}&SOM &Analyzed TCP data traffic patterns using SOM and detected anomalies based on abnormal behavior. \\
		Lichodzijewski et al. \cite{lichodzijewski2002host} &SOM &Applied SOM to host based intrusion detection.\\
		Lichodzijewski et al. \cite{lichodzijewski2002dynamic}&SOM &Applied a hierarchical NN to detect intruders, emphasizing on the development of relational hierarchies and time representation.\\
		Amini et al. \cite{amini2006rt}&SOM \& ART &Applied SOM combined with ART networks in real-time IDS. \\
		Depren et al. \cite{depren2005intelligent}&SOM \& J.48 Decision Tree &Applied SOM combined with J.48 decision tree algorithm in IDS to detect anomaly and misuses intelligently. \\
		Golovko et al. \cite{golovko2006neural}& Multi-Layer Perceptrons (MLP) & Presented a two-tier IDS architecture. PCA in the first tier reduces input dimensions, while MLP in the second tier detects and recognizes attacks with low detection time and high accuracy.\\

		\midrule
		
		\textit{\textbf{\underline{Data Clustering}}} & & \\\\
		Leung et al. \cite{leung2005unsupervised} &Density \& Grid Based Clustering &Applied an unsupervised clustering strategy in density and grid based clustering algorithms to detect anomalies.\\
		Chimphlee et al. \cite{chimphlee2006anomaly} &Fuzzy Rough Clustering &Applied the idea of Fuzzy set theory and fuzzy rough C-means clustering algorithms in IDS to detect abnormal behaviors in networks, producing excellent results.\\
		Jianliang et al. \cite{jianliang2009application} &K-Means &Applied K-means clustering in IDS to detect intrusions and anomalies.\\
		Muniyandi et al. \cite{muniyandi2012network} &K-Means with C4.5 Decision Trees &Applied K-means clustering combined with C4.5 decision tree models to detect intrusive and anomalous behavior in networks and systems.\\
		Casas et al. \cite{casas2012unsupervised} &Sub-space Clustering &Implemented a unique unsupervised outliers and anomaly detection approach using Sub-Space Clustering and Multiple Evidence Accumulation techniques to exactly identify different kinds of network intrusions and attacks such as DoS/DDoS, probing attacks, buffer overflows, etc.\\
		Zanero et al. \cite{zanero2004unsupervised} &Two-Tier Clustering &Applied a novel bi-layered clustering technique, in which the first layer constitutes of clustering of packets and the second layer is responsible for anomaly detection and time correlation, to detect intrusions. \\
		Gaddam et al. \cite{gaddam2007k}&K-Means \& ID3 Decision Trees &Applied K-means clustering combined with ID3 decision tree models to detect intrusive and anomalous behavior in systems.\\
		
		Zhong et al. \cite{zhong2007clustering} &Centroid Based Clustering &Presented a survey on intrusion detection techniques based on centroid clustering as well as other popular unsupervised approaches.\\
		
		Greggio et al. \cite{greggio2017anomaly} &Finite GMM &An unsupervised greedy learning of finite GMM is used for anomaly detection in intrusion detection system. \\
		
		\midrule
		\textit{\textbf{\underline{Blind Signal Separation}}} & & \\\\
		 	Xu et al. \cite{xu2005adaptive}&PCA &Applied PCA and SVM in IDS. \\
		 	Wang et al. \cite{wang2006identifying}&PCA &Applied a novel approach to translate each network connection into a data vector, and then applied PCA to reduce its dimensionality and detect anomalies. \\
		 	Golovko et al. \cite{golovko2007dimensionality}&PCA &Applied PCA and dimensionality reduction techniques in attack recognition and anomaly detection.\\
		 	Guan et al. \cite{guan2009fast}&NMF &Applied NMF algorithms to capture intrusion and network anomalies. \\ 
		   \hline
		\end{tabular}
	}
\end{table*}
       

\subsection{Anomaly/Intrusion Detection}
The increasing use of networks in every domain has increased the risk of network intrusions, which makes user privacy and the security of critical data vulnerable to attacks. According to the annual computer crime and security survey 2005 \cite{gordon20052005}, conducted by the combined teams of CSI (Computer Security Institute) and FBI (Federal Bureau of Investigation), total financial losses faced by companies due to the security attacks and the network intrusions were  estimated as US \$130 million. Moreover, according to Symantec Internet Security Threat Report \cite{reportSTR}, approximately 5000 new vulnerabilities were identified in the year 2015. In addition, more than 400 million new variants of malware and 9 major breaches were detected exposing 10 million identities. Therefore, insecurity in today's networking environment has given rise to the ever-evolving domain of network security and intrusion/anomaly detection \cite{reportSTR}.
       
In general, Intrusion Detection Systems (IDS) recognize or identify any act of security breach within a computer or a network; specifically, all requests which could compromise the confidentiality and availability of data or resources of a system or a particular network. Generally, intrusion detection systems can be categorized into three types: (1) signature-based intrusion detection systems; (2) anomaly detection systems; and (3) compound/hybrid detection systems, which include selective attributes of both preceding systems.
      
Signature detection, also known as misuse detection, is a technique that was initially used for tracing and identifying misuses of user's important data, computer resources, and intrusions in the network based on the previously collected or stored signatures of intrusion attempts. The most important benefit of a signature-based system is that a computer administrator can exactly identify the type of attack a computer is currently experiencing based on the sequence of the packets defined by stored signatures. However, it is nearly impossible to maintain the signature database of all evolving possible attacks, thus this pitfall of the signature-based technique has given rise to anomaly detection systems.

Anomaly Detection System (ADS) is a modern intrusion and anomaly detection system. Initially, it creates a baseline image of a system profile, its network and user program activity. Then, on the basis of this baseline image, ADS classifies any activity deviating from this behavior as an intrusion. Few benefits of this technique are: firstly, they are capable of detecting insider attacks such as using system resources through another user profile; secondly, each ADS is based on a customized user profile which makes it very difficult for attackers to ascertain which types of attacks would not set an alarm; and lastly, it detects unknown behavior in a computer system rather than detecting intrusions, thus it is capable of detecting any unknown sophisticated attack which is different from the users' usual behavior. However, these benefits come with a trade-off, in which the process of training a system on a user's `normal' profile and maintaining those profiles is a time consuming and challenging task. If an inappropriate user profile is created, it can result in poor performance. Since ADS detects any behavior that does not align with a user's normal profile, its false alarm rate can be high. Lastly, another pitfall of ADS is that a malicious user can train ADS gradually to accept inappropriate traffic as normal.  
        %
         

As anomaly and intrusion detection has been a popular research area since the origin of networking and Internet, numerous supervised as well as unsupervised \cite{tsai2009intrusion} learning techniques have been applied to efficiently detect intrusions and malicious activities. However, latest research focuses on the application of unsupervised learning techniques in this area due to the challenge and promise of using big data for optimizing networks.
         
Initial work focuses on the application of basic unsupervised clustering algorithms for detecting intrusions and anomalies. In 2005, an unsupervised approach was proposed based on density and grid based clustering to accurately classify the high-dimensional dataset in a set of clusters; those points which do not fall in any cluster are marked as abnormal \cite{leung2005unsupervised}. This approach has produced good results but false positive rate was very high. In a follow-up work, another improved approach that used fuzzy rough C-means clustering was introduced \cite{chimphlee2006anomaly} \cite{zhong2007clustering}. K-means clustering is also another famous approach used for detecting anomalies which was later proposed in 2009 \cite{jianliang2009application}, which showed great accuracy and outperformed existing unsupervised methods. However, later in 2012, an improved method which used K-means clustering combined with C4.5 decision tree algorithm was proposed \cite{muniyandi2012network} to produce more efficient results than prior approaches. \cite{lin2015cann} combines cluster centers and nearest neighbors for effective feature representation which ensures a better intrusion detection, a limitation with this approach is that it is not able to detect user to resource and remote to local attacks. Another scheme using unsupervised learning approach for anomaly detection is presented in \cite{mazel2015hunting}. The presented scheme combines subspace clustering and correlation analysis to detect anomalies and provide protection against unknown anomalies; this experiment used WIDE backbone networks data \cite{sony2000traffic} spanning over six years and produced better results then previous K-means based techniques. Work presented in \cite{papalexakis2014network} shows that for different intrusions schemes, there are a small set of measurements required to differentiate between normal and anomalous traffic; the authors used two co-clustering schemes to perform clustering and to determine which measurement subset contributed the most towards accurate detection.
         
Another famous approach for increasing detection accuracy is ensemble learning, work presented in \cite{mivskovic2014application} employed many hybrid incremental ML approach with gradient boosting and ensemble learning to achieve better detection performance. Authors in \cite{haq2015application} surveyed anomaly detection research from 2009 to 2014 and find out the a unique algorithmic similarity for anomaly detection in Internet traffic: most of the algorithms studied have following similarities 1) Removal of redundant information in training phase to ensure better learning performance 2) Feature selection usually performed using unsupervised techniques and increases the accuracy of detection 3) Use ensembles classifiers or hybrid classifiers rather than baseline algorithms to get better results. Authors in \cite{hamalainen2014artificial} have developed an artifical immune system based intrusion detection system they have used density based spatial clustering of applications with noise to develop an immune system against the network intrusion detection.
         
The application of unsupervised intrusion detection in cloud network is presented in \cite{chaturvedi2016study} where authors have proposed a fuzzy clustering ANN to detect the less frequent attacks and improve the detection stability in cloud networks. Another application of unsupervised intrusion detection system for clouds is surveyed in \cite{modi2013survey}, where fuzzy logic based intrusion detection system using supervised and unsupervised ANN is proposed for intrusion detection; this approach is used for DOS and DDoS attacks where the scale of the attack is very large. Network intrusion anomaly detection (NIDS) based on K-means clustering are surveyed in \cite{weller2015survey};  this survey is unique as it provides distance and similarity measure of the intrusion detection and this perspective has not been studied before 2015. Unsupervised learning based application of anomaly detection schemes for wireless personal area networks, wireless sensor networks, cyber physical systems, and WLANs is surveyed in \cite{mitchell2014survey}.
         
Another paper \cite{ahmed2016survey} reviewing  anomaly detection has presented the application of unsupervised SVM and clustering based applications in network intrusion detection systems. Unsupervised discretization algorithm is used in Bayesian network classifier for intrusion detection, which is based on Bayesian model averaging \cite{xiao2014bayesian}; the authors show that the proposed algorithm performs better than the Na{\"\i}ve Bayes classifier in terms of accuracy on the NSL-KDD intrusion detection dataset. Border gateway protocol (BGP)---the core Internet inter-autonomous systems (inter-AS) routing protocol---is also error prone to intrusions and anomalies. To detect these BGP anomalies, many supervised and unsupervised ML solutions (such as hidden Markov models and principal component analysis) have been proposed in literature \cite{al2017bgp} for anomaly and intrusion detection. Another problem for anomaly detection is low volume attacks, which have become a big challenge for network traffic anomaly detection. While long range dependencies (LRD) are used to identify these low volume attacks, LRD usually works on aggregated traffic volume; but since the volume of traffic is low, the attacks can pass undetected. To accurately identify low volume abnormalities, Assadhan et al. \cite{assadhan2017anomaly} proposed the examination of LRD behavior of control plane and data plane separately to identify low volume attacks.    
       
Other than clustering, another widely used unsupervised technique for detecting malicious and abnormal behavior in networks is SOMs. The specialty of SOMs is that they can automatically organize a variety of inputs and deduce patterns among themselves, and subsequently determine whether the new input fits in the deduced pattern or not, thus detecting abnormal inputs  \cite{rhodes2000multiple} \cite{kayacik2003capability}. SOMs have also been used in host-based intrusion detection systems in which intruders and abusers are identified at a host system through incoming data traffic \cite{lichodzijewski2002dynamic}, later on a more robust and efficient technique was proposed to analyze data patterns in TCP traffic \cite{zanero2005analyzing}. Furthermore, complex NNs have also been applied to solve the same problem and remarkable results have been produced. A few examples include the application of ART combined with SOM \cite{amini2006rt}. The use of PCA can also be seen in detecting intrusions \cite{wang2006identifying}. NMF has also been used for detecting intruders and abusers \cite{guan2009fast}, and lastly dimension reduction techniques have also been applied to eradicate intrusions and anomalies in the system \cite{golovko2007dimensionality}. For more applications, refer to Table \ref{Table:ANIDS Systems}, which classifies different network anomaly and intrusion detection systems on the basis of unsupervised learning techniques discussed earlier.

\subsection{Network Operations, Optimizations and Analytics}

Network management comprises of all the operations included in initializing, monitoring and managing of a computer network based on its network functions, which are the primary requirements of the network operations. The general purpose of network management and monitoring systems is to ensure that basic network functions are fulfilled, and if there is any malfunctioning in the network, it should be reported and addressed accordingly. Following is a summary of different network optimization tasks achieved through unsupervised learning models.
    
               \begin{table*}
               \caption{Unsupervised Learning Techniques employed for Network Operations, Optimizations and Analytics \label{Table:NNs in WSNs}}
    {
	\centering
		\scriptsize
	\begin{tabular}{ p{3cm} p{2.5cm} p{8cm} p{2.5cm} }

		\toprule
		Reference &  Technique & Brief Summary & Network Type\\
		\midrule
		
\multicolumn{3}{l}{\textbf{\emph{\underline{Hierarchical Representations/ Deep Learning}}}}\\
\\
		Kulakov et al. \cite{kulakov2005application} &ART fuzzy &Applied ART NNs at clusterheads and sensor nodes to extract regular patterns, reducing data for lesser communication overhead. &WSN\\
		
		Akojwar et al. \cite{akojwar2008improving} &ART   &Applied ART at each network node for data aggregation. &WSN\\
		
    	Li et al. \cite{li2014distributed}  &DNN &Applied different DNN layers corresponding to WSN layers in order to compress data. &WSN\\
		
		Gelenbe et al. \cite{gelenbe2002cognitive} &RNN &Applied RNN to achieve optimal QoS in cognitive packet networks. &Cognitive networks\\
   
		Cordina et al. \cite{cordina2008increasing} &SOM  &Applied SOM to cluster nodes into categories based on node location, energy and concentration; some nodes becomes clusterheads. &WSN\\
		
		Enami et al. \cite{enami2010energy} &SOM  &Applied SOM to categorize and select nodes with higher energy levels to become clusterheads based on node energy levels. &WSN\\
		
		Dehni et al. \cite{dehni2006power} &SOM & Applied SOM followed by K-means to cluster and select clusterheads in WSNs. &WSN\\
		
		Oldewurtel et al. \cite{oldewurtel2006neural} &SOM  &Applied SOMs in clusterheads to find patterns in data. &WSN\\
				
		Barreto et al. \cite{barreto2005condition} &DNN &Applied a competitive neural algorithm for condition monitoring and fault detection in 3G cellular networks. &Cellular networks\\
				
		Moustapha et al. \cite{moustapha2008wireless} &RNN  &Applied RNN for fault detection. RNN, which is deployed in each sensor node, takes inputs from neighboring nodes, and generates outputs for comparison with the generated data; if the difference exceeds a certain threshold, the node is regarded as anomalous. & WSN \\
		
	    \midrule
	    \textit{\textbf{\underline{Data Clustering}}} & & \\
	    \\
		Hoan et al. \cite{hoang2010fuzzy}  &Fuzzy C-Means Clustering &Applied fuzzy C-means clustering technique to select nodes with the highest residual energy to gather data and send information using an energy-efficient routing in WSNs. &WSN \\
		
		Oyman et al. \cite{oyman2004multiple} &K-Means Clustering &Applied K-means clustering to design multiple sink nodes in WSNs. &WSN \\
		
		Zhang et al. \cite{zhang2006trust} &K-Means Partitioning &Applied K-means clustering to identify compromised nodes and applied Kullback-Leibler (KL) distance to determine the trustworthiness (reputation) of each node in a trust-based WSN. &WSN \\
		     
		\midrule
	    \textit{\textbf{\underline{Blind Signal Separation}}} & & \\
\\
	    Kapoor et al. \cite{kapoor2015outlier}&PCA&Applied PCA to resolve the problem of cooperative spectrum sensing in cognitive radio networks. &Cognitive radio networks \\
	    
	    Ristaniemi et al. \cite{ristaniemi2002advanced}&ICA&Applied ICA based CDMA receivers to separate and identify mixed source signals.&CDMA \\
	    
		Ahmed et al. \cite{ahmed2005probabilistic} &PCA & Applied PCA to evaluate the degree of confidence in detection probability provided by a WSN. The probabilistic approach is a deviation from the idealistic assumption of sensing coverage used in a binary detection model. &WSN \\
		
		Chatzigiannakis et al. \cite{chatzigiannakis2006hierarchical} &PCA &Applied PCA for hierarchical anomaly detection in a distributed WSN. &WSN\\
        \\
		\hline
		
	\end{tabular}
	}
\end{table*}

\vspace{2mm}
\subsubsection{QoS/ QoE Optimization}
        
QoS and QoE are measures of the service performance and end-user experience, respectively. QoS mainly deals with the performance as seen by the user being measured quantitatively, while QoE is a qualitative measure of a subjective metrics experienced by the user. QoS/QoE for Internet services (especially multimedia content delivery services) is crucial in order to maximize the user experience. With the dynamic and bursty nature of Internet traffic, computer networks should be able to adapt to these changes without compromising the end-user experiences. As QoE is quite subjective, it heavily relies on the underlying QoS which is affected by different network parameters; \cite{mushtaq2012empirical} and \cite{alreshoodi2013survey} suggested different measurable factors to determine the overall approximation of QoS such as error rates, bit rate, throughput, transmission delay, availability, jitter, etc. Furthermore, these factors are used to correlate QoS with QoE in the perspective of video streaming where QoE is essential to end-users.
        
The dynamic nature of Internet dictates network design for different applications to maximize QoS/QoE, since there is no predefined adaptive algorithm that can be used to fulfill all the necessary requirements for prospective application. Due to this fact, ML approaches are employed in order to adapt to the real-time network conditions and take measures to stabilize/maximize the user experience. \cite{testolin2014machine} employed a hybrid architecture having unsupervised feature learning with supervised classification for QoE-based video admission control and resource management. Unsupervised feature learning in this system is carried out by using a fully connected NN comprising RBMs, which capture descriptive features of video that are later classified by using a supervised classifier. Similarly, \cite{przylucki2011assessment} presents an algorithm to estimate the Mean Opinion Score, a metric for measuring QoE, for VoIP services by using SOM to map quality metrics to features.

Moreover, research has shown that QoE-driven content optimization leads to the optimal utilization of network. Ahammad et al. \cite{ahammad2014qoe} showed that 43\% of the bit overhead on average can be reduced per image delivered on the web. This is achieved by using the quality metric VoQS (Variation of Quality Signature), which can arbitrarily compare two images in terms of web delivery performance. By applying this metric for unsupervised clustering of large image dataset, multiple coherent groups are formed in device-targeted and content-dependent manner. In another study \cite{mocanu2014deep}, deep learning is used to assess the QoE of 3D images that have yet to show good results compared with the other deterministic algorithms. The outcome is a Reduced Reference QoE assessment process for automatic image assessment, and it has a significant potential to be extended to work on 3D video assessment. 

In \cite{Bernardo2009RL}, a unique technique of the model-based RL approach is applied to improve bandwidth availability, and hence throughput performance, of a network. The MRL model is embedded in a node that creates a model of the operating environment, and uses it to generate virtual states and rewards for the virtual actions taken. As the agent does not need to wait for the real states and rewards from the operating environment, it can explore various kinds of actions on the virtual operating environment within a short period of time which helps to expedite the learning process, and hence the convergence rate to the optimal action. In \cite{Liang2010RL}, a MARL approach is applied in which nodes exchange Q-values among themselves and select their respective next-hop nodes with the best possible channel conditions while forwarding packets towards the destination. This helps to improve throughput performance as nodes in a network ensure that packets are successfully sent to the destination in a collaborative manner.



\vspace{2mm}
\subsubsection{TCP Optimization}

Transmission Control Protocol (TCP) is the core end-to-end protocol in TCP/IP stack that provides reliable, ordered and error-free delivery of messages between two communicating hosts. Due to the fact that TCP provides reliable and in-order delivery, congestion control is one of the major concerns of this protocol, which is commonly dealt with the algorithms defined in \textit{RFC 5681}. However, classical congestion control algorithms are sub-optimal in hybrid wired/wireless networks as they react to packet loss in the same manner in all network situations. In order to overcome this shortcoming of classical TCP congestion control algorithms, an ML-based approach is proposed in \cite{geurts2004machine}, which employs a supervised classifier based on features learned for classifying a packet loss due to congestion or link errors. Other approaches to this problem currently employed in literature includes using RL that uses fuzzy logic based reward evaluator based on game theory \cite{hwang2005cooperative}. Another promising approach, named \textit{Remy} \cite{winstein2013tcp}, uses a modified model of \textit{Markov decision process} based on three factors: 1) prior knowledge about the network; 2) a traffic model based on user needs (i.e., throughput and delay); and 3) an objective function that is to be maximized. By this learning approach, a customized best-suited congestion control scheme is produced specifically for that part of the network, adapted to its unique requirements. However, classifying packet losses using unsupervised learning methods is still an open research problem and there is a need of real-time adaptive congestion control mechanism for multi-modal hybrid networks. 
    
For more applications, refer to Table \ref{Table:NNs in WSNs}, which classifies different various network optimization and operation works on the basis of their network type and the unsupervised learning technique used.

     \begin{table*}[!ht]	
           	\caption{Dimensionality Reduction Techniques employed for Networking Applications   \label{Table:Dimensionality Reduction Techniques}}
    {
           \centering
		\scriptsize
	\begin{tabular}{p{3cm} p{3cm} p{8cm} p{2.5cm}}
		\toprule
		Reference & Technique & Brief Summary & Network/Technology Type   \\
		\midrule

    O'Shea et al. \cite{o2017introduction} &Autoencoders & Applied autoencoders to design an end-to-end communication system that can jointly learn  transmitter  and  receiver  implementations  as  well  as signal  encodings in unsupervised manner. &MIMO \\
    
    O'Shea et al. \cite{o2017deep} &Autoencoders &A new approach for designing and optimizing the physical layer is explored using autoencoders for dimensionality reduction. &MIMO \\
     
    O'Shea et al. \cite{o2016unsupervised} &Convolutional Autoencoders &Applied autoencoders for representation learning of structured radio communication signals. &Software Radio/ Cognitive Radio \\
    
    Huang et al. \cite{huang2016new} &Multi-dimensional Scaling &Applied distance based subspace dimensionality reduction technique for anomaly detection in data traffic. &Internet Traffic \\
    
    Zoha et al. \cite{zoha2016learning} &Multi-dimensional Scaling &Used MDS to preprocess a statistical dataset for cell outage detection in SON. &SON \\
    
    Shirazinia et al. \cite{shirazinia2015power} &Sparse Gaussian Method &Applied sparse Gaussian method for linear dimensionality reduction over noisy channels in wireless sensor networks. &WSN \\
    
    Hou et al. \cite{hou2011svm} &PCA &Linear and nonlinear dimensionality reduction techniques along with support vector machine has be experimentally tested for cognitive radio. &Cognitive Radio \\
    
    Khalid et al. \cite{khalid2015network} &PCA &Applied L1 norm PCA for dimensionality reduction in network intrusion detection system. &Internet Traffic \\
    
    Goodman et al. \cite{goodman2015using} &PCA &Applied PCA for diemensionality reduction in anomaly detection for cyber security applications. &SMS \\
    
    Patwari et al. \cite{patwari2005manifold} & Manifold Learning &Proposed a manifold learning based visualization tool for network traffic visualization and anomaly detection. &Internet Traffic \\
    
    Lopez et al. \cite{lopez2018deep} & Transfer Learning and t-SNE &Used transfer learning for multimedia web mining and t-SNE for dimensionality reduction and visualization of web mining resultant model. &Multimedia Web \\
    
    Ban et al. \cite{ban2016towards} & Clustering and t-SNE &Proposed an early threat detection scheme using darknet data, where clustering is used for threat detection and dimensionality reduction for visualization is performed by using t-SNE. &Internet Traffic \\
    
\midrule
	
	\end{tabular}
	}
\end{table*}

\subsection{Dimensionality Reduction \& Visualization} 
\label{DRapplications}

Network data usually consists of multiple dimensions. To apply machine learning techniques effectively the number of variables are needed to be reduced. Dimensionality reduction schemes have a number of significant potential applications in networks. In particular, dimensionality reduction can be used to facilitate network operations (e.g., for anomaly/intrusion detection, reliability analysis, or for fault prediction) and network management (e.g., through visualization of high-dimensional networking data). A tabulated summary of various research works using dimensionality reduction techniques for various kinds of networking applications is provided in Table \ref{Table:Dimensionality Reduction Techniques}. 

Dimensionality reduction techniques have been used to improve the effectiveness of the anomaly/intrusion detection system. Niyaz et al. \cite{niyaz2016deep} proposed a DDoS detection system in SDN where dimensionality reduction is used for feature extraction and reduction in unsupervised manner using stacked sparse autoencoders. Cordero et al. \cite{cordero2016analyzing} proposed a flow based anomaly intrusion detection using replicator neural network. Proposed network is based on an encoder and decoder where the hidden layer between encoder and decoder performs the dimensionality reduction in unsupervised manner, this process also corresponds to PCA. Similarly Chen et al. \cite{chen2017novel} have proposed another anomaly detection procedure where dimensionality reduction for feature extraction is performed using multi-scale PCA and then using wavelet analysis, so that the anomalous traffic is separated from the flow. Dimensionality reduction using robust PCA based on minimum covariance determinant estimator for anomaly detection is presented in \cite{matsuda2017traffic}. Thaseen et al. \cite{thaseen2014intrusion} applied PCA for dimensionality reduction in network intrusion detection application. To improve the performance of intrusion detection scheme, another algorithm based on dimensionality reduction for new feature learning using PCA is presented in \cite{subba2016enhancing} \cite{muttaqien2016increasing}. Almusallam et al. \cite{almusallam2017dimensionality} have reviewed the dimensionality reduction schemes for intrusion detection in multimedia traffic and proposed an unsupervised feature selection scheme based on the dimensionality reduced multimedia data. 

Dimensionality reduction using autoencoders performs a vital role in fault prediction and reliability analysis of the cellular networks, this work also recommends deep belief networks and autoencoders as logical fault prediction techniques for self organizing networks \cite{kumar2017fault}. Most of the Internet applications use encrypted traffic for communication, previously deep packet inspection (DPI) was considered a standard way of classifying network traffic but with the varying nature of the network application and randomization of port numbers and payload size DPI has lost its significance. Authors in \cite{nascimento2014multi} have proposed a hybrid scheme for network traffic classification. Proposed scheme uses extreme machine learning, genetic algorithms and dimensionality reduction for feature selection and traffic classification. Ansari et al. \cite{ansari2015fuzzy} applied fuzzy set theoretic approach for dimensionality reduction along with fuzzy C-mean clustering algorithm for quality of web usage. In another work, Alsheikh et al. \cite{alsheikh2015toward} used Shrinking Sparse AutoEncoders (SSAE) for representing high-dimensional data and utilized SSAE in compressive sensing settings. 

Visualization of high dimensional data in lower dimension representation is another application of dimensionality reduction. There are many relevant techniques such as PCA and t-SNE that can be used to extract the underlying structure of high-dimensional data, which can then be visualized to aid human insight seeking and decision-making \cite{maaten2008visualizing}. A number of researchers have proposed to utilize dimensionality reduction techniques to aid visualization of networking data. Patwari et al. \cite{patwari2005manifold} proposed a manifold learning based visualization tool for network traffic visualization and anomaly detection. Labib et al. \cite{labib2006application} proposed a PCA-based for the detection and visualization of networking attacks in which PCA is used for the dimensionality reduction of the feature vector extracted from KDD network traffic dataset. Lokovc et al. \cite{lokovc2016k} used t-SNE for depicting malware fingerprints in their proposed network intrusion detection system. Ancona et al. \cite{ancona2006visualizing} proposed a rectangular dualization scheme for visualizing the underlying network topology. Cherubin et al. \cite{cherubin2015conformal} used dimensionality reduction and t-SNE of clustering and visualization of botnet traffic. Finally, a lightweight platform for home Internet monitoring is presented in \cite{marshlightweight} where PCA and t-SNE is used for dimensionality reduction and visualization of the network traffic. A number of tools are readily available---e.g., Divvy \cite{lewis2013divvy}, Weka \cite{holmes1994weka}---that implement dimensionality reduction and other unsupervised ML techniques (such as PCA and manifold learning) and allow exploratory data analysis and visualization of high-dimensional data.

Dimensionality reduction techniques and tools have been utilized in all kinds of networks and we present some recent examples related to self-organizing networks (SONs) and software defined radios (SDRs). Liao et al. \cite{liao2015network} proposed a semi supervised learning scheme for anomaly detection in SON based on dimensionality reduction and fuzzy classification technique. Chernov et al. \cite{chernov2015location} used minor component analysis (MCA) for dimensionality reduction as a preprocessing step for user level statistical data in LTE-A networks to detect the cell outage. Zoha et al. \cite{zoha2016learning} used multi-dimensional scaling (MDS), a dimensionality reduction scheme, as part of the preprocessing step for cell outage detection in SON. Another data driven approach by Zoha et al. \cite{zoha2015data} also uses MDS for getting low dimensional embedding of target key point indicator vector as a preprocessing step to automatically detect cell outage in SON. Turkka et al. \cite{turkka2012approach} used PCA for dimensionality reduction of drive test samples to detect cell outages autonomously in SON. Conventional routing schemes are not sufficient for the fifth generation of communication systems. Kato et al. \cite{kato2017deep} proposed a supervised deep learning based routing scheme for heterogeneous network traffic control. Although supervised approach performed well, but gathering a lot of heterogeneous traffic with labels, and then processing them with a plain ANN is computationally extensive and prone to errors due to the imbalanced nature of the input data and the potential for overfitting. In 2017, Mao et al. \cite{mao2017routing} has presented a deep learning based approach for routing and cost effective packet processing. The proposed model uses deep belief architecture and benefits from the dimensionality reduction property of restricted Boltzmann machine. The proposed work also provides a novel Graphics Processing Unit (GPU) based router architecture. The detailed analysis shows that deep learning based SDR and routing technique can meet the changing network requirements and massive network traffic growth. The routing scheme proposed in \cite{mao2017routing} outperforms conventional open shortest path first (OSPF) routing technique in terms of throughput and average delay per hop. 

\subsection{Emerging Networking Applications of Unsupervised Learning}  

Next generation network architectures such as Software defined Networks (SDN), Self Organizing Networks (SON), and Internet of Things (IoT) are expected to be the basis of future intelligent, adaptive, and dynamic networks \cite{qadir2014sdns}. ML techniques will be at the center of this revolution providing aforementioned properties. This subsection covers the recent applications of unsupervised ML techniques in SDNs, SONs, and IoTs.
   
\vspace{2mm}
\subsubsection{Software Defined Networks}

SDN is a disruptive new networking architecture that simplifies network operating and managing tasks and provides infrastructural support for novel innovations by making the network programmable \cite{nunes2014survey}. In simple terms, the idea of programmable networks is to simply decouple the data forwarding plane and control/decision plane, which is rather tightly coupled in current infrastructure. The use of SDN can also be seen in managing and optimizing networks as network operators go through a lot of hassle to implement high level security policies in term of distributed low level system configurations, thus SDN resolves this issue by decoupling the planes and giving network operators better control and visibility over network, enabling them to make frequent changes to network state and providing support for high-level specification language for network control \cite{kim2013improving}. SDN is applicable in a wide variety of areas ranging from enterprise networks, data centers, infrastructure based wireless access networks, optical networks to home and small businesses, each providing many future research opportunities \cite{nunes2014survey}. 

Unsupervised ML techniques are seeing a surging interest in SDN community as can be seen by a spate of recent work. A popular application of unsupervised ML techniques in SDNs relates to the application of \textit{intrusion detection and mitigation of security attacks} \cite{ashraf2014handling}. Another approach for detecting anomalies in cloud environment using unsupervised learning model has been proposed by Dean et al. \cite{dean2012ubl} that uses SOM to capture emergent system behavior and predict unknown and novel anomalies without any prior training or configuration. A DDoS detection system for SDN is presented in \cite{niyaz2016deep} where stacked autoencoders are used to detect DDoS attacks. A density peak based clustering algorithm for DDoS attack is proposed as a new method to review the potentials of using SDN to develop an efficient anomaly detection method \cite{he2017software}. Goswami et al. \cite{goswami2017intelligent} have recently presented an intelligent threat aware response system for SDN using reinforcement learning, this work also recommends using unsupervised feature learning to improve the threat detection process. Another framework for anomaly detection, classification, and mitigation for SDN is presented in \cite{da2016atlantic} where unsupervised learning is used for \textit{traffic feature analysis}. Zhang et al. \cite{zhang2017sdnforensics}  have presented a forensic framework for SDN and recommended K-means clustering for anomaly detection in SDN.  Another work \cite{amaral2016machine} discusses the potential opportunities for using unsupervised learning for \textit{traffic classification} in SDN. Moreover, deep learning and distributed processing can also be applied to such models in order to better adapt with evolving networks and contribute to the future of SDN infrastructure as a service.

\vspace{2mm}
\subsubsection{Self Organizing Networks}

Self organizing networks (SON) is another new and popular research regime in networking, SON are inspired from the biological system which works in self organization and achieve the task by learning from the surrounding environment. As the connected network devices are growing exponentially, and the communication cell size has reduced to femtocells, the property of self organization is becoming increasingly desirable \cite{aliu2013survey}. A reinforcement learning based approach for designing self organization based small cell network is presented in \cite{bennis2013self}. Feasibility of SON application in fifth generation  (5G) of wireless communication is studied in \cite{imran2014challenges} and the study shows that without (supervised as well as unsupervised) ML support, SON is not possible. Application of ML techniques in SON has become a very important research area as it involves learning from the surroundings for intelligent decision-making and reliable communication \cite{latif2017artificial}.
     
Application of different ML-based SON for heterogeneous networks is considered in \cite{wang2015artificial}, this paper also describes the unsupervised ANN, hidden Markov models and reinforcement learning techniques employed for better learning from the surroundings and adapting accordingly. PCA and clustering are the two mostly used unsupervised learning schemes utilized for parameter optimization and feature learning in SON where as reinforcement learning, fuzzy reinforcement learning, Q learning, double Q learning and deep reinforcement learning are the major schemes used for interacting with the environment \cite{aliu2013survey}. These ML schemes are used in self-configuration, self-healing, and self-optimization schemes. Game theory is another unsupervised learning approach used for designing self optimization and greedy self configuration design of SON systems \cite{misra2017self}. Authors in \cite{zhang2014swarm} proposed an unsupervised ANN for link quality estimation of SON which outperformed simple moving average and exponentially weighted moving averages.
     
\vspace{2mm}
\subsubsection{Internet of Things}

Internet of things (IoT) is an emerging paradigm with a growing academia and industry interest. IoT is new networking paradigm and it is expected to be deployed in health care, smart cities, industry, home automation, agriculture, and industry. With such a vast plane of applications, IoT needs ML to collect and analyze data to make intelligent decisions. The key challenge that IoT must deal with is the extremely large scale (billions of devices) of future IoT deployments \cite{wen2017fog}. Designing, analyzing and predicting are the three major tasks and all involves ML, a few examples of unsupervised ML are shared next. Gubbi et al. \cite{gubbi2013internet} recommend using unsupervised ML techniques for feature extraction and supervised learning for classification and predictions. Given the scale of the IoT, a large amount of data is expected in the network and therefore requires a load balancing method, a load balancing algorithm based on restricted Boltzmann machine is proposed in \cite{kim2017load}. Online clustering scheme form dynamic IoT data streams is described in \cite{puschmann2017adaptive}. Another work describing an ML application in IoT recommends a combination of PCA and regression for IoT to get better prediction \cite{assem2016machine}. Usage of clustering technique in embedded systems for IoT applications is presented in \cite{lee2016integrating}. An application using denoising autoencoders for acoustic modeling in IoT is presented in \cite{lin2017multi}. 

\subsection{Lessons Learnt}
Key leassons drawn from the review of unsupervised learning in networking applications are summarized below: 

\begin{enumerate}

\item A recommended and well studied method for unsupervised Internet traffic classification in literature is data clustering combined with the latent representation learning on traffic feature set by using autoencoders. Min-max ensemble learning will help to increase the efficiency of unsupervised learning if required.
    
\item Semi supervised learning is also an appropriate method for Internet traffic classification given some labeled traffic data and channel characteristics are available for initial model training. 
    
\item Application of generative models and transfer learning for the Internet traffic classification has not been explored properly in literature and can be a potential research direction.  

\item The overwhelming growth in network traffic and expected surge in traffic with the evolution of 5G and IoT also elevates the level of threat and anomalies in network traffic. To deal with these anomalies in Internet traffic, data clustering, PCA, SOM, and ART are well explored unsupervised learning techniques in literature. Self-taught learning has also been explored as a potential solution for anomaly detection and remains a possible research direction for future research in anomaly detection in network traffic. 
    
\item Unsupervised learning techniques for network management and optimization is a very less explored area as compared to anomaly detection and traffic classification. Applications of NN, RBM, Q learning, and deep reinforcement learning techniques to Internet traffic for management and optimization is an open research area. 
    
\item Current state of the art in dimensionality reduction in network traffic is based on PCA and multidimensional scaling. Autoencoders, t-SNE, and manifold learning is a potential area of research in terms of dimensionality reduction and visualization.

    

\end{enumerate}
    
\section{Future Work: Some Research Challenges and Opportunities}
\label{futurework}
	    
This section provides a discussion on some open directions for future work and the relevant opportunities in applying unsupervised ML in the field of networking. 

\subsection{Simplified Network Management}

While new network architectures such as SDN have been proposed in recent years to simply network management, network operators are still expected to know too much, and to correlate between what they know about how their network is designed with the current network's condition through their monitoring sources. Operators who manage these requirements by wrestling with complexity manually will definitely welcome any respite that they can get from (semi-)automated unsupervised machine learning. As highlighted in by \cite{sommer2010outside}, for ML to become pervasive in networking, the ``semantic gap''---which refers to the key challenge of transferring ML results into actionable insights and reports for the network operator---must be overcome. This can facilitate a shift from a \emph{reactive} interaction style for network management, where the network manager is expected to check maps and graphs when things go wrong, to a \emph{proactive} one, where automated reports and notifications are created for different services and network regions. Ideally, this would be abstract yet informative, such as Google Maps Directions, e.g. ``there is heavier traffic than usual on your route'' as well as suggestions about possible actions. This could be coupled with automated correlation of different reports coming from different parts of the network. This will require a move beyond mere notifications and visualizations to more substantial synthesis through which potential sources of problems can be identified. Another example relates to making measurements more user-oriented. Most users would be more interested in QoE instead of QoS, i.e., how the current condition of the network affects their applications and services rather than just raw QoS metrics. The development of measurement objectives should be from a business-eyeball perspective---and not only through presenting statistics gathered through various tools and protocols such as traceroute, ping, BGP, etc. with the burden of putting the various pieces of knowledge together being on the user. 

\subsection{Semi-Supervised Learning for Computer Networks}

Semi-supervised learning lies between supervised and unsupervised learning. The idea behind semi-supervised learning is to improve the learning ability by using unlabeled data incorporation with small set of labeled examples. In computer networks, semi-supervised learning is partially used in anomaly detection and traffic classification and has great potential to be used with deep unsupervised learning architectures like generative adversarial networks for improving the state of the art in anomaly detection and traffic classification. Similarly user behavior learning for cyber security can also be tackled in a semi-supervised fashion. A semi-supervised learning based anomaly detection approach is presented in \cite{ashfaq2017fuzziness}. The presented approach used large amounts of unlabeled samples together with labeled samples to build a better intrusion detection classifier. In particular, a single hidden layer feed-forward NN is trained to output a fuzzy membership vector. The results show that using unlabeled samples help significantly improve the classifier's performance. In another work, Watkins et al. \cite{watkins2017using} have proposed a semi-supervised learning with 97\% accuracy to filter out non-malicious data in millions of queries that Domain Name Service (DNS) servers receive. 

\subsection{Transfer Learning in Computer Networks}
    
Transfer learning is an emerging ML technique in which knowledge learned from one problem is applied to a different but related problem \cite{pan2010survey}. Although it is often thought that for ML algorithms, the training and future data must be in the same feature space and must have same distribution, this is not necessarily the case in many real-world applications. In such cases, it is desirable to have \textit{transfer learning}, or knowledge transfer between the different task domains. Transfer learning has been successfully applied in computer vision and NLP applications but its implementation for networking has not been witnessed---even though in principle, this can be useful in networking as well due to the similar nature of Internet traffic and enterprise network traffic in many respects. Bacstuug et al. \cite{bacstuug2015transfer} used transfer learning based caching procedure for wireless networks providing backhaul offloading in 5G networks. 


\subsection{Federated Learning in Computer Networks}

Federated learning is a collaborative ML technique, which does not make use of centralized training data, and works by distributing the processing on different machines. Federated learning is considered to be the next big thing in cloud networks as they ensure privacy of the user data and less computation on the cloud to reduce the cost and energy \cite{konevcny2016federated}. System and method for network address management in federated cloud is presented in \cite{gokhale2017system} and application of federated IoT and cloud computing for health care is presented in \cite{abawajy2017federated}. An end-to-end security architecture for federated cloud and IoT is presented in \cite{massonet2017end}. 
   
\subsection{General Adversarial Networks (GANs) in Computer Networks}

Adversarial networks---based on generative adversarial network (GAN) training originally proposed by Goodfellow and colleagues at the University of Montreal \cite{goodfellow2014generative}---have recently emerged as a new technique using which machines can be trained to predict outcomes by only the observing the world (without necessarily being provided labeled data). An adversarial network has two NN models:  a generator—which is responsible for generating some type of data from some random input—and a discriminator, which has the task of distinguishing between input from the generator or a real data set. The two NNs optimize themselves together resulting in more realistic generation of data by the generator, and a better sense of what is plausible in the real world for the discriminator. The use of GANs for ML in networking can improve the performance of ML-based networking applications such as anomaly detection in which malicious users have an incentive to adversarial craft new attacks to avoid detection by network managers. 

\section{Pitfalls and Caveats of Using Unsupervised ML in Networking}
\label{pitfalls}
	    
With the benefits and intriguing results of unsupervised learning, there also exists many shortcomings that are not addressed widely in the literature. Some potential pitfalls and caveats related to unsupervised learning are discussed next.

\subsection{Inappropriate Technique Selection} To start with, the first potential pitfall could be the selection of technique. Different unsupervised learning and predicting techniques may have excellent results on some applications while performing poorly on others---it is important to choose the best technique for the task at hand. Another reason could be a poor selection of features or parameters on which basis predictions are made---thus parameter optimization is also important for unsupervised algorithms. 
	    
\subsection{Lack of Interpretability of Some Unsupervised ML Algorithms} Some unsupervised algorithms such as deep NNs operate as a blackbox, which makes it difficult to explain and interpret the working of such models. This makes the use of such techniques unsuitable for applications in which interpretability is important. As pointed out in \cite{sommer2010outside}, understandability of the semantics of the decisions made by ML is especially important for the operational success of ML in large-scale operational networks and its acceptance by operators, network managers, and users. But prediction accuracy and simplicity are often in conflict \cite{breiman2001statistical}. As an example, the greater accuracy of NNs accrues from its complex nature in which input variables are combined in a nonlinear fashion to build a complicated hard-to-explain model; with NNs it may not be possible to get interpretability as well since they make a tradeoff in which they sacrifice interpretability to achieve high accuracy. There are various ongoing research efforts that are focused on making techniques such as NNs less opaque \cite{sturm2016interpretable}. Apart from the focus on NNs, there is a general interest in making AI and ML more explainable and interpretable---e.g., the Defense Advanced Research Projects Agency or DARPA's \textit{explainable AI project}\footnote{\url{https://www.darpa.mil/program/explainable-artificial-intelligence}} is aiming to develop explainable AI models (leveraging various design options spanning the performance-vs-explainability tradeoff space) that can explain the rationale of their decision-making so that users are able to appropriately trust these models particularly for new envisioned control applications in which optimization decisions are made autonomously by algorithms. 
	   
\subsection{Lack of Operational Success of ML in Networking}

In literature, researchers have noted that despite substantial academic research, and practical applications of unsupervised learning in other fields, we see that there is a dearth of practical applications of ML solutions in operational networks---particular for applications such as network intrusion detection \cite{sommer2010outside}, which are challenging problems for a number of reasons including 1) the very high cost of errors; 2) the lack of training data; 3) the semantic gap between results and their operational interpretation; 4) enormous variability in input data; and finally, 5) fundamental difficulties in conducting sound performance evaluations. Even for other applications, the success of ML and its wide adoption in practical systems at scale lags the success of ML solutions in many other domains. 

\subsection{Ignoring Simple Non-Machine-Learning Based Tools}

One should also keep in mind a common pitfall that academic researchers may suffer from: which is not realize that network operators may have have simpler non-machine learning based solutions that may work as well as na{\"\i}ve ML based solutions in practical settings. Failure to examine the ground realities of operational networks will undermine the effectiveness of ML based solutions. We should expect ML based solutions to augment and supplement rather than replace other non-machine-learning based solutions---at least for the foreseeable future. 

\subsection{Overfitting} 
\begin{figure}
    	\begin{center}
    	 	\includegraphics[width=0.4\textwidth]{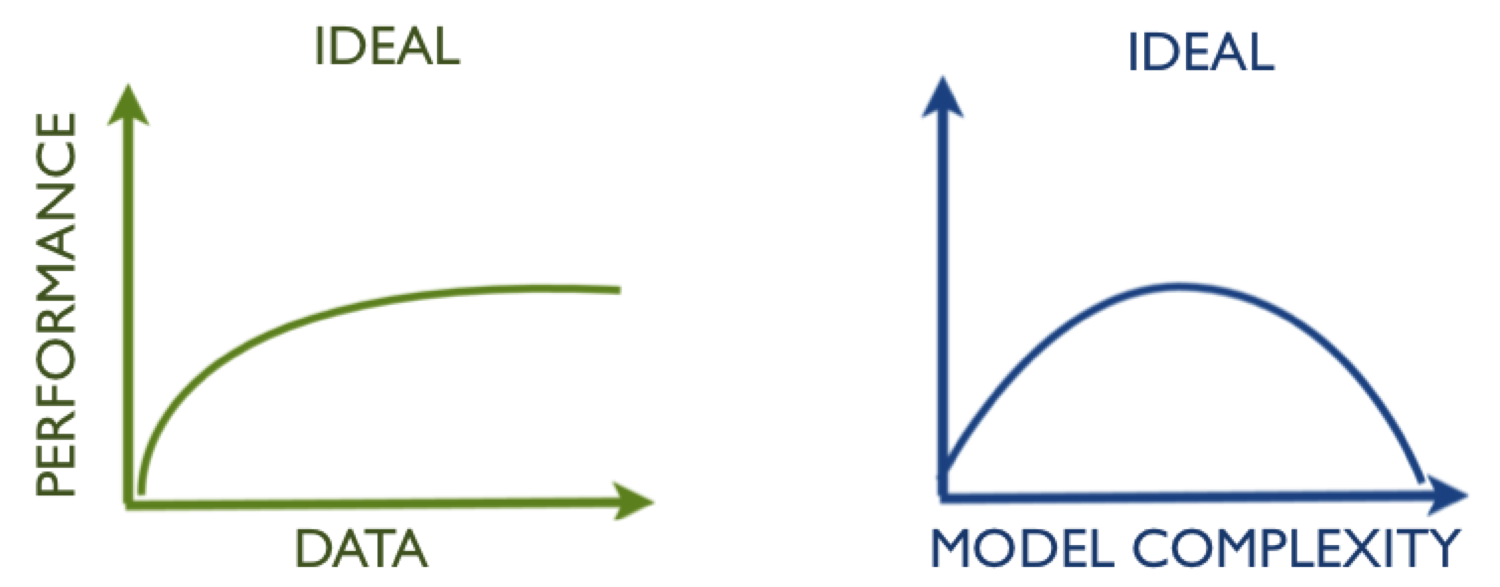}
    		 \caption{Intuitively, we expect the ML model's performance to improve with more data but to deteriorate in performance if the model becomes overly complex for the data. Figure adapted from \protect\cite{zhu2016we}.}
     	 	\label{fig:bigdata_modelcomplexity}
      	\end{center}
     \end{figure}

Another potential issue with unsupervised models is overfitting; it corresponds to a model representing the noise or random error rather than learning the actual pattern in data. While commonly associated with supervised ML, the problem of overfitting lurks whenever we learn from data and thus is applicable to unsupervised ML as well. As illustrated in Figure \ref{fig:bigdata_modelcomplexity}, ideally speaking, we expect ML algorithms to provide improved performance with more data; but with increasing model complexity, performance starts to deteriorate after a certain point---although, it is possible to get poorer results empirically with increasing data when working with unoptimized out-of-the-box ML algorithms \cite{zhu2016we}. According to the Occam Razor principle, the model complexity should be commensurate with the amount of data available, and with overly complex models, the ability to predict and generalize diminishes. Two major reasons of overfitting could be the overly large size of learning model and less sample data used for training purposes. Generally data is divided into two portions (actual data and stochastic noise); due to the unavailability of labels or related information, unsupervised learning model can overfit the data, which causes issues in testing and deployment phase. Cross validation, regularization, and Chi-squared testing are highly recommended designing or tweaking an unsupervised learning algorithm to avoid overfitting \cite{domingos2012few}.

\subsection{Data Quality Issues} It should be noted that all ML is data dependent, and the performance of ML algorithms is affected largely by the nature, volume, quality, and representation of data. Data quality issues must be carefully considered since any problem with the data quality will seriously mar the performance of ML algorithms. A potential problem is that dataset may be \textit{imbalanced} if the samples size from one class is very much smaller or larger than the other classes \cite{amin2016comparing}. In such imbalanced datasets, the algorithm must be careful not to ignore the rare class by assuming it to be noise. Although, imbalanced datasets are more of a nuisance for supervised learning techniques, they may also pose problems for unsupervised and semi-supervised learning techniques. 

\subsection{Inaccurate Model Building} It is difficult to build accurate and generic models since each model is optimized for certain kind of applications. Unsupervised ML models should be applied after carefully studying the application and the suitability of the algorithm in such settings \cite{zhang2007avoiding}. For example, we highlight certain issues related to the unsupervised task of clustering: 1) random initialization in K-means is not recommended; 2) number of clusters are not known before the clustering operation as we do not have labels; 3) in the case of hierarchical clustering, we don not know when to stop and this can cause increase in the time complexity of the process; and 4) evaluating the clustering result is very tricky since the ground truth is mostly unknown. 
    
\subsection{Machine Learning in Adversarial Environments} Many networking problems, such as anomaly detection, is an adversarial problem in which the malicious intruder is continually trying to outwit the network administrators (and the tools used by the network administrators). In such settings, machine learning that learns from historical data may not perform due to clever crafting of attacks specifically for circumventing any schemes based on previous data. 

\vspace{4mm}
    
Due to these challenges, pitfalls, and weaknesses, due care must be exercised while using unsupervised and semi-supervised ML. These pitfalls can be avoided in part by using various best practices \cite{ng2011advice}, such as end-to-end learning pipeline testing, visualization of learning algorithm, regularization, proper feature engineering, dropout, sanity checks through human inspection---whichever is appropriate for the problem's context.

\section{Conclusions}
\label{conclusions}
	
We have provided a comprehensive survey of machine learning tasks and latest unsupervised learning techniques and trends along with a detailed discussion of the applications of these techniques in networking related tasks. Despite the recent wave of success of unsupervised learning, there is a scarcity of unsupervised learning literature for computer networking applications, which this survey aims to address. The few previously published survey papers differ from our work in their focus, scope, and breadth; we have written this paper in a manner that carefully synthesizes the insights from these survey papers while also providing contemporary coverage of recent advances. Due to the versatility and evolving nature of computer networks, it was impossible to cover each and every application; however, an attempt has been made to cover all the major networking applications of unsupervised learning and the relevant techniques. We have also presented concise future work and open research areas in the field of networking, which are related to unsupervised learning, coupled with a brief discussion of significant pitfalls and challenges in using unsupervised machine learning in networks.


\bibliographystyle{ieeetr}
\bibliography{UnsupervisedLearning}

\end{document}